\documentclass[floats,floatfix,showpacs,amssymb,amsmath,prd,onecolumn,superscriptaddress,nofootinbib, aps]{revtex4-2}
\usepackage[utf8]{inputenc}
\usepackage{hyperref, amsmath, amsfonts}
\usepackage{graphicx}
\usepackage{aas_macros}
\hypersetup{colorlinks,bookmarksnumbered,
citecolor=cyan, urlcolor=magenta}

\providecommand{\adsurl}[1]{\href{#1}{ADS}}
\usepackage{natbib}
\usepackage[usenames, dvipsnames]{xcolor}
\usepackage{booktabs}
\usepackage{ulem}
\usepackage{tikz}
\usepackage{esvect}
\usetikzlibrary{"arrows", "automata", "backgrounds", "calendar", "chains", "matrix", "mindmap", "patterns", "petri", "shadows", "shapes.geometric", "shapes.misc", "spy", "trees"}
\usetikzlibrary{arrows,shapes}
\usetikzlibrary{trees}
\usetikzlibrary{matrix,arrows} 				% For commutative diagram
\usetikzlibrary{positioning}				% For "above of=" commands
\usetikzlibrary{calc,through}				% For coordinates
\usetikzlibrary{decorations.pathreplacing}  % For curly braces
\usepackage{pgffor}							% For repeating patterns

\newcommand{\e}[1]{\times 10^{#1}}

\begin{document}
\hspace{5.2in} \mbox{CALT-TH/2022-011}

\title{Atom Interferometer Tests of Dark Matter}
\author{Yufeng Du}
\affiliation{Walter Burke Institute for Theoretical Physics, California Institute of Technology, Pasadena, CA 91125, USA}
\author{Clara Murgui}
\affiliation{Walter Burke Institute for Theoretical Physics, California Institute of Technology, Pasadena, CA 91125, USA}
\author{Kris Pardo}
% \email{kpardo@caltech.edu}
\affiliation{Jet Propulsion Laboratory, California Institute of Technology, Pasadena, CA 91109, USA}
\affiliation{Walter Burke Institute for Theoretical Physics, California Institute of Technology, Pasadena, CA 91125, USA}
\author{Yikun Wang}
\affiliation{Walter Burke Institute for Theoretical Physics, California Institute of Technology, Pasadena, CA 91125, USA}
\author{Kathryn M. Zurek}
\affiliation{Walter Burke Institute for Theoretical Physics, California Institute of Technology, Pasadena, CA 91125, USA}

\date{\today}

\begin{abstract}
    Direct detection experiments for dark matter are increasingly ruling out large parameter spaces. However, light dark matter models with particle masses $<$ GeV are still largely unconstrained. Here we examine a proposal to use atom interferometers to detect a light dark matter subcomponent at sub-GeV masses. We describe the decoherence and phase shifts caused by dark matter scattering off of one ``arm" of an atom interferometer using a generalized dark matter direct detection framework. This allows us to consider multiple channels: nuclear recoils, hidden photon processes, and axion interactions. We apply this framework to several proposed atom interferometer experiments. Because atom interferometers are sensitive to extremely low momentum deposition and their coherent atoms may give them a boost in sensitivity, these experiments will be highly competitive and complementary to other direct detection methods. In particular, atom interferometers are uniquely able to probe a dark matter sub-component with $m_\chi \lesssim 10~\rm{keV}$. We find that, for a mediator mass $m_\phi=10^{-5}m_\chi$, future atom interferometers could close a gap in the existing constraints on nuclear recoils down to $\bar{\sigma}_n \sim 10^{-42}~\rm{cm}^2$ for $m_\chi \sim 10^{-5} - 10^{-1}~\rm{MeV}$ dark matter masses.
\end{abstract}

\maketitle

\section{Introduction}

Although we have seen dark matter (DM) through its astrophysical and cosmological effects \citep{Zwicky1933, Rubin1978, Rubin1980, Ostriker1974, Tyson1990, Wittman2000, Spergel2003, Planck2018}, we have yet to directly detect DM. The traditional direct detection methods, which rely on measuring nuclear recoils, have succeeded in placing ever stronger limits on Weakly Interacting Massive Particles (WIMPs), but have not led to any detections \citep{Ahlen1987, Aprile2018, Cui2017, Akerib2017, PandaX}. In the last several years, motivated by theories of dark sectors \cite{Boehm:2003hm,Strassler:2006im,Pospelov:2007mp,Hooper:2008im,Feng:2008ya,Arkani-Hamed:2008hhe,Zurek:2008qg,Hall:2009bx,Lin:2011gj,Hochberg:2014dra}, many new types of direct detection channels and experiments have been proposed to probe lighter DM masses. Electronic transitions in a variety of materials \citep{Essig2012, Essig2012b, Graham2012, Lee2015, Essig2015, Hochberg2016, Hochberg2016b, Hochberg2016c, Hochberg2017, Derenzo2017, Hochberg2017b, Bloch2017, Essig2017, Kadribasic2018, Hochberg2018, Kurinsky2019, Heikinheimo2019, Emken2019, Coskuner2019b, Geilhufe2020, Griffin2020}, molecular excitations \citep{Essig2017b, Arvanitaki2018b, Essig2019}, and phonon excitations \citep{Schutz2016, Knapen2017b, Knapen2018, Griffin2018, Acanfora2019} have been identified as promising new directions.

Recently, there has been an interest in using quantum sensors, and especially those in space, to detect dark matter \citep[\textit{e.g.},][]{Carney2021, Belenchia2022}. Atom interferometers, in particular, offer an interesting and complementary pathway to dark matter detection compared to conventional searches. These experiments use falling atoms to measure differential forces along two paths, and they can be used as accelerometers \citep[\textit{e.g.},][]{Kasevich1991, Geiger2020}. Recent suggestions for DM detection with atom interferometers have focused on ultralight DM with couplings to either the photon or the electron, which would induce temporal oscillations in fundamental constants. Although these couplings were first investigated for their possible detection with atomic clock experiments \citep{Arvanitaki:2014faa, Stadnik:2015kia, Hees:2016gop, https://doi.org/10.48550/arxiv.2005.14694}, these couplings would also be measurable in an atom interferometer via induced phase shifts \citep{Geraci2016, Graham2016, Arvanitaki2018}. Notably, these tests would only be applicable to specific types of atom interferometers -- those that rely on atomic transitions and shared lasers.

Another proposal considers the decoherence caused by DM scattering in atom interferometers \citep{Riedel2013, Riedel2017}. This would be applicable to all types of atom interferometers. Previous work has shown that future atom interferometers would provide promising constraints on DM that interacts with the Standard Model (SM) through nuclear recoils \citep{Riedel2017}. Atom interferometers have two useful features that enable these constraints: the coherence of the atoms may allow for an $N^2$ enhancement in the scattering rate, and they do not have a minimum energy threshold. This means that atom interferometers can probe to extremely low momentum depositions, and thus low dark matter masses.

In this paper, we consider observables that potentially get coherent enhancements: phase-shifts for cold atomic clouds, and both contrast loss and phase shifts for matter interferometers\footnote{Recent work~\cite{Badurina:2024nge} shows that atom interferometers based on diffuse atom clouds do not receive $N^2$ enhancements to the decoherence rate when one-body measurements are performed. Experiments based on Bose–Einstein condensates (BECs) exhibit a similar behavior than diffuse atomic clouds and we will therefore treat them as such. We leave a comprehensive treatment for future work.}, and expand the atom interferometer detection mechanism to further scattering channels. Fig.~\ref{fig:cartoon} gives a schematic description of the detection strategy. We standardize and generalize the formalism of Ref.~\citep{Riedel2017} using the methods of Ref.~\citep{Trickle2020}. We then consider nuclear recoils, hidden photon processes, and coherent axion scattering. Our main results are summarized in Fig.~\ref{fig:results}, which shows that atom interferometers can be uniquely powerful for probing nuclear recoil of DM with masses $m_{\chi} \lesssim 1~\rm{MeV}$. Although the constraints through the other channels are not competitive, we include rough analytic estimates.

The paper is organized as follows. In Section~\ref{sec:overview} we discuss the overall method, explain our general formalism, and describe the atom interferometer experiments we consider in this paper. We then apply our formalism in Section~\ref{sec:dm_rates} to nuclear recoils (\ref{sec:nuc_recoils}), and dark photon processes and coherent axion scattering (\ref{sec:other_process}). We discuss how these atom interferometer constraints compare to other relevant constraints in Section~\ref{sec:other_constraints}, and we conclude in Section~\ref{sec:conclusion}.

Throughout this paper, we assume natural units, $\hbar = c = 1$ and consider a DM particle, $\chi$, that is a subcomponent of the total DM in the Universe: $\Omega_{\chi} = 0.05~\Omega_{\rm{DM}}$. Light DM with $m_{\chi} \lesssim 1~\rm{MeV}$ interacting with nucleons at a large enough rate to be detectable in an atom interferometer will typically come into thermal equilibrium at some point in the history of the Universe, giving rise to a variety of constraints, discussed in detail in Ref.~\cite{Knapen:2017}; considering a dark matter subcomponent evades these bounds. The code used for our calculations and figures is publicly available at: \url{https://github.com/kpardo/atom_interferometer_dm}.

\section{Direct Detection with Atom Interferometers}\label{sec:overview}

In this section, we review the effects of DM scattering on atom interferometers. We develop the general formula that describes the scattering rate measurable by an atom interferometer, and then discuss the specific atom interferometer experiments we consider in the paper.

\begin{figure*}
    \centering
\includegraphics[width=0.99\textwidth
    %, trim={0 4cm 0 4cm},clip
    ]{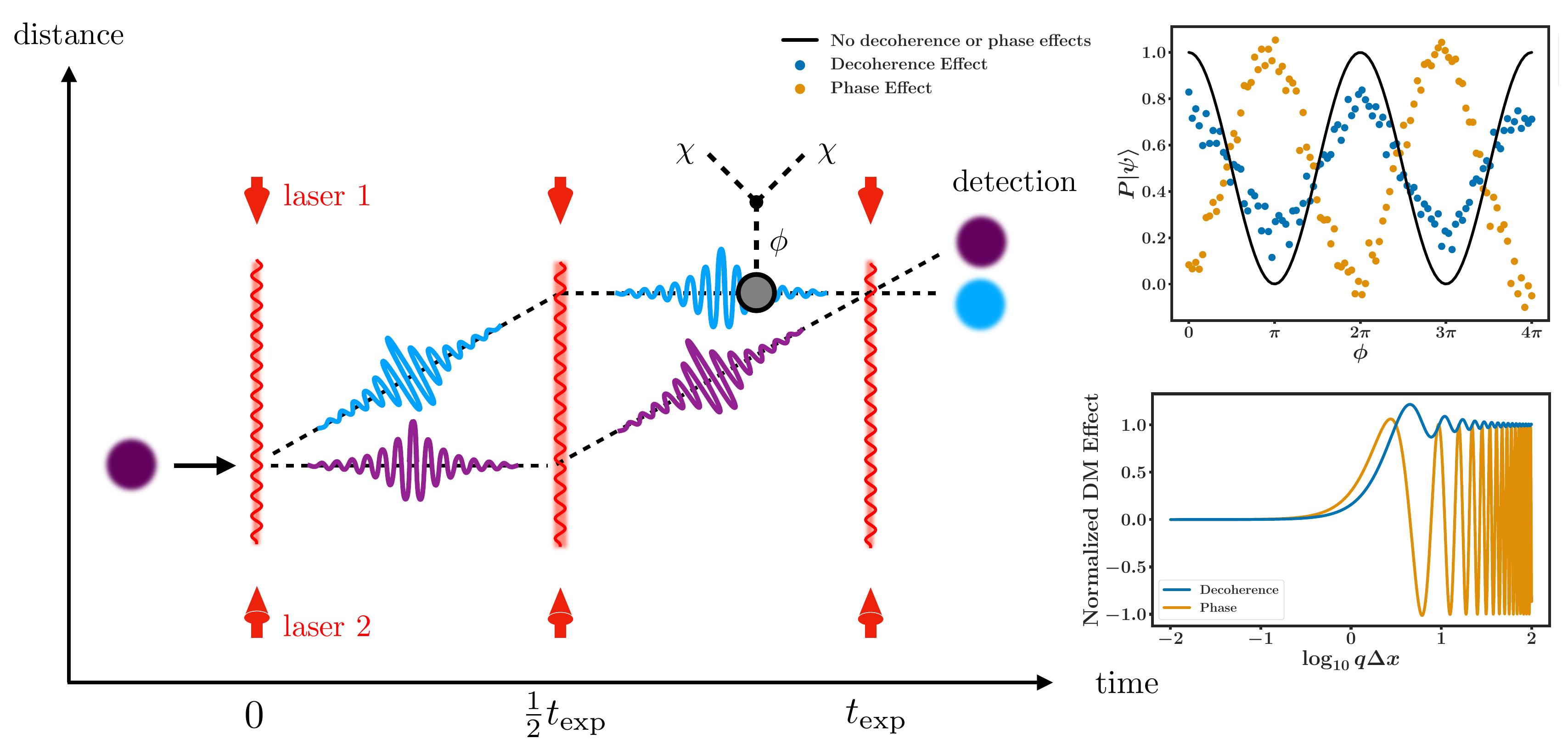}
    \caption{\textit{Left:} Cartoon of dark matter detection with atom interferometry. Three laser pulses are used to separate, redirect, and recombine the atom wavepackets. Atom interferometers are a promising avenue for detecting DM because they are sensitive to low momentum transfers and the coherence of the atom clouds may allow for an enhancement in the DM scattering rates. \textit{Right, Top:} Cartoon of expected fringe measurement (\textit{e.g.}, probability of the state $|\psi\rangle$ at the end of the atom interferometer sequence as a function of final laser phase) for the main observables from DM scattering. The black line gives the expected fringe if there are no DM or other effects. The blue (orange) points give simulated data points that show the effect of decoherence (phase shifts) on the fringe. \textit{Right, Bottom:} Observable effects as a function of the DM momentum transfer ($q$) and the spatial distance between the wavepacket paths of the atoms ($\Delta x$). The blue line shows the decoherence induced by DM scattering (see Eqn.~\ref{eqn:decoherence_effect}), and the orange line shows the phase shift (see Eqn.~\ref{eqn:phase_effect}). Both are maximally affected by DM at $q\sim 1/\Delta x$, which is where we expect the DM to ``resolve" the clouds.}
    \label{fig:cartoon}
\end{figure*}

In essence, DM scattering will cause decoherence and phase shifts in an atom interferometer (see Fig.~\ref{fig:cartoon}). Consider a Mach-Zender atom interferometer \citep[for a general review of atom interferometers, see Ref.][]{Geiger2020}. Analogously to the conventional Mach-Zender interferometer, here the de-Broglie wave nature of the atoms plays the role of light, while laser pulses act as beamsplitters and mirrors. The atoms start in a coherent wavepacket, then are split by a laser pulse into two different wavepackets that travel along two different paths. Two more laser pulses reverse the motion of these wavepackets and recombine them. The state of the recombined wavepacket is then read, with the relative populations depending on the initial prepared state, the timing of the laser pulses, and any differential forces felt along the different paths.

The main observables from an atom interferometer are the visibility, $V$, and phase, $\phi$. The visibility is a measure of the decoherence of the system, and the phase measures the path differences between the two arms. Previously, atom interferometer measurements would require calibration of the fringe over several measurements with varying laser phase to give a proper measurement of either the phase or visibility \citep[see, e.g.,][]{Kasevich1991}. However, modern atom interferometers are capable of measuring a spatial fringe, and thus $V$, with each measurement \citep[some ways of doing this are shown in][]{Dickerson2013, Sugarbaker2013}. Unless the phase shifts associated with systematic effects are known exactly, $\phi$ cannot be measured from a single spatial fringe without prior knowledge. However, the relative phase between two different measurements or between two atom interferometers on the same laser can be measured \cite[e.g.,][]{Foster2002, Chiow2009}.

We will assume that both the phase and visibility can be reliably measured, and we will consider the case where the only relevant force causing decoherence comes from scattering with DM particles, which can be considered to be a thermal bath. The first laser pulse splits the population of the atoms and prepares the system $|\Psi \rangle$ in a quantum superposition of two states: the ``blue" and the ``purple" wavepackets, following the same color coding as in Fig.~\ref{fig:cartoon}. In particular, any DM scattering with either the ``purple" or ``blue" wavepacket will cause decoherence in the final state. The density of states in the \{purple, blue\} basis is given by:
\begin{equation}
    \rho = \frac{1}{2} \left(\begin{matrix} 1 & \gamma \\
    \gamma^* & 1\end{matrix}\right) \; ,
\end{equation}
where $\gamma$ is the decoherence factor, which contains both the decoherence and phase effects. Specifically, this factor can be decomposed into a dimensionless decoherence, $s$ (real, positive), and a phase, $\phi$ (real), via: $\gamma = \exp (-s+i\phi)$.

When the wavepackets recombine, the evolved state $|\tilde \Psi \rangle$ is measured in the basis of the initial state $| \Psi \rangle$ with a probability of
\begin{equation}
    P(|\tilde \Psi \rangle , |\Psi \rangle) = \text{Tr}\{\rho \, |\Psi \rangle \langle \Psi | \} = \frac{1}{2}(1+\mathrm{Re}(\gamma)) = \frac{1}{2}(1+e^{-s}\cos\phi) \; .
\end{equation}
The visibility of an experiment, $V = e^{-s}$, is given by measuring the amplitude of the fringe. As explained above, the absolute phase cannot be measured -- instead a differential phase must be measured by comparing two measurements at different times or measurements of different atom interferometers on the same laser. DM scattering will affect \textit{both} of these observables.

The decoherence factor will be directly related to the number of DM scattering events and how its resolution (de Broglie wavelength) compares to the separation between the two interferometer modes. Let $R$ give the rate of dark matter scattering per target mass that leads to decoherence or phase shifts in the atom interferometer. Then, over a measurement time, $t_{\rm{exp}}$, the decoherence factor accumulated will be:
\begin{equation}
    \gamma = \exp \left[ - \frac{m_T}{N_{\rm ind}} \int_0^{t_{\rm{exp}}} R~dt \right] \; ,
\end{equation}
where $m_T$ gives the target mass -- the mass of the atom cloud in the interferometer, and $N_{\rm{ind}}$ is the number of {\it independent} objects that are measured at the read-out port. For an unentangled atom interferometer (\textit{e.g.,} a cold atom interferometer, such as the BECCAL, GDM, Stanford, and AEDGE experiments considered in this paper) with $N_A$ independent atoms, $N_{\rm ind} = N_A$. For entangled atom interferometers, matter interferometers (such as the MAQRO and Pino experiments considered in this paper), and quantum resonators, only one independent measurement is made per shot -- the quantum state of the macroscopic object. Thus, $N_{\rm ind}=1$.

Now, we turn our attention to calculating the rate per target mass, $R$. Consider a non-relativistic DM particle with mass $m_{\chi}$ that has some spin-independent scattering interaction with SM nucleons, mediated by a field of mass $m_{\phi}$. For reasons that will be fully discussed later, we stress that the observables we focus on in this paper are only effective for {\textbf{spin-independent}} interactions. The process occurs as shown in the Feynman diagram below, with the corresponding reference cross-section $\bar \sigma$,
\begin{equation}
\begin{gathered}
    \includegraphics[width=0.17\textwidth]{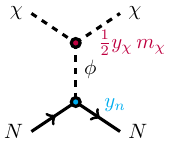}
    \end{gathered} : \quad \bar \sigma =\frac{y_\chi^2 \, y_n^2}{4\pi}\frac{\mu^2}{(m_\chi^2 v_0^2+m_\phi^2)^2} ,
    \label{eqn:RefXSec}
\end{equation}
where $y_i$ are the dimensionless coupling factors, $\mu = m_\chi m_N/(m_{\chi} + m_N)$ is the reduced mass of the DM and each nucleon, and $v_0=230~\rm{km/s}$ is the local average velocity of the DM. When necessary, we set $y_\chi = \sqrt{4\pi}$ in this paper.

As we will see, atom interferometers have the best constraining power at low $m_{\chi}$. We will be most interested in the limit $m_{\chi} \ll m_N$. Thus, we replace the reduced mass with $\mu \approx m_{\chi}$ throughout. In addition, we will assume, without loss of generality, that the DM $\chi$ is a real scalar. Other scenarios can be adapted to the reference cross section from Eqn.~\ref{eqn:RefXSec} by redefining the coupling $y_\chi$.

We would like to find the rate of DM scattering events that lead to decoherence or phase shifts in an atom interferometer. The differential rate per target mass is given by \cite[see, \textit{e.g.,}][]{Trickle2020}:
\begin{equation}
    \frac{dR}{d\omega} = \frac{1}{\rho_T}\frac{\rho_\chi}{m_\chi} \int d^3\mathbf{v}~f(\mathbf{v}) \frac{d\Gamma}{d\omega}\; ,
\end{equation}
where $\rho_T$ is the density of the target material, $\omega$ is the energy deposition, $\rho_\chi$ is the DM density, and $f(\mathbf{v})$ is the DM velocity distribution function. The differential interaction rate, $d\Gamma/d\omega$, can be written using the formalism of Ref~\cite{Trickle2020} as:
\begin{equation}
    \frac{d\Gamma}{d\omega} = \frac{\pi \bar{\sigma}}{\mu^2} \int \frac{d^3{\bf q}}{(2\pi)^3} \mathcal{F}_{\rm{med}}^2(\mathbf{q}) S(\mathbf{q}, \omega) \delta(\omega - \omega_{\mathbf{q}}) \; ,
\end{equation}
where $\mathbf{q}$ is the momentum deposition, $\bar{\sigma}$ is a reference cross section, $\mathcal{F}_{\rm{med}}(\mathbf{q})$ is the mediator form factor, and $S(\mathbf{q}, \omega_{\mathbf{q}})$ is the dynamic structure factor. It is this last factor that encodes atom interferometer-specific physics.

The dynamic structure factor encodes the response of a detector to a given momentum deposition \citep[see, e.g., Eqn.~14 of][]{Trickle2020}. It includes terms that conserve energy in the target and account for a momentum transfer dependence in the target material. In addition, the dynamic structure factor should encode the restrictions of an atom interferometer as a DM detector. Namely, only scattering events that produce momentum transfer along the atom interferometer separation direction, $\mathbf{\Delta x}$, produce decoherence and phase shifts. The probability that a scattering event will contribute to the decoherence factor of a single atom is \citep{Joos1985, Hornberger2003, Riedel2013, Riedel2017}:
\begin{equation} \label{eq:pdec}
    p_{\rm{decoh}} = 1-\exp\left[i\mathbf{q}\cdot \mathbf{\Delta x}\right] \; .
\end{equation}

To get a better idea of this effect, let us consider the real and imaginary parts separately. If we take only the real component of $p_{\rm{decoh}}$ and integrate over the angle between $\mathbf{q}$ and $\mathbf{\Delta x}$, we find:
\begin{equation}\label{eqn:decoherence_effect}
    \frac{1}{2}\int_{-1}^{1} \mathrm{Re}(p_{\rm{decoh}}) ~d\cos\theta_{\mathbf{q\Delta x}} = 1 -\mathrm{sinc}(q\Delta x) \; ,
\end{equation}
where $q\equiv \|\mathbf{q}\|$, $\Delta x\equiv \|\mathbf{\Delta x}\|$, and the $\frac{1}{2}$ factor later cancels with other angular integrals. This is the ``decoherence effect" plotted in Fig.~\ref{fig:cartoon}. Note that the decoherence is maximal when the DM ``resolves'' the clouds, $q \sim 1/\Delta x$, and quickly vanishes for $q < 1/\Delta x$.

\begin{figure*}[ht]
    \centering
    \includegraphics[width=\textwidth]{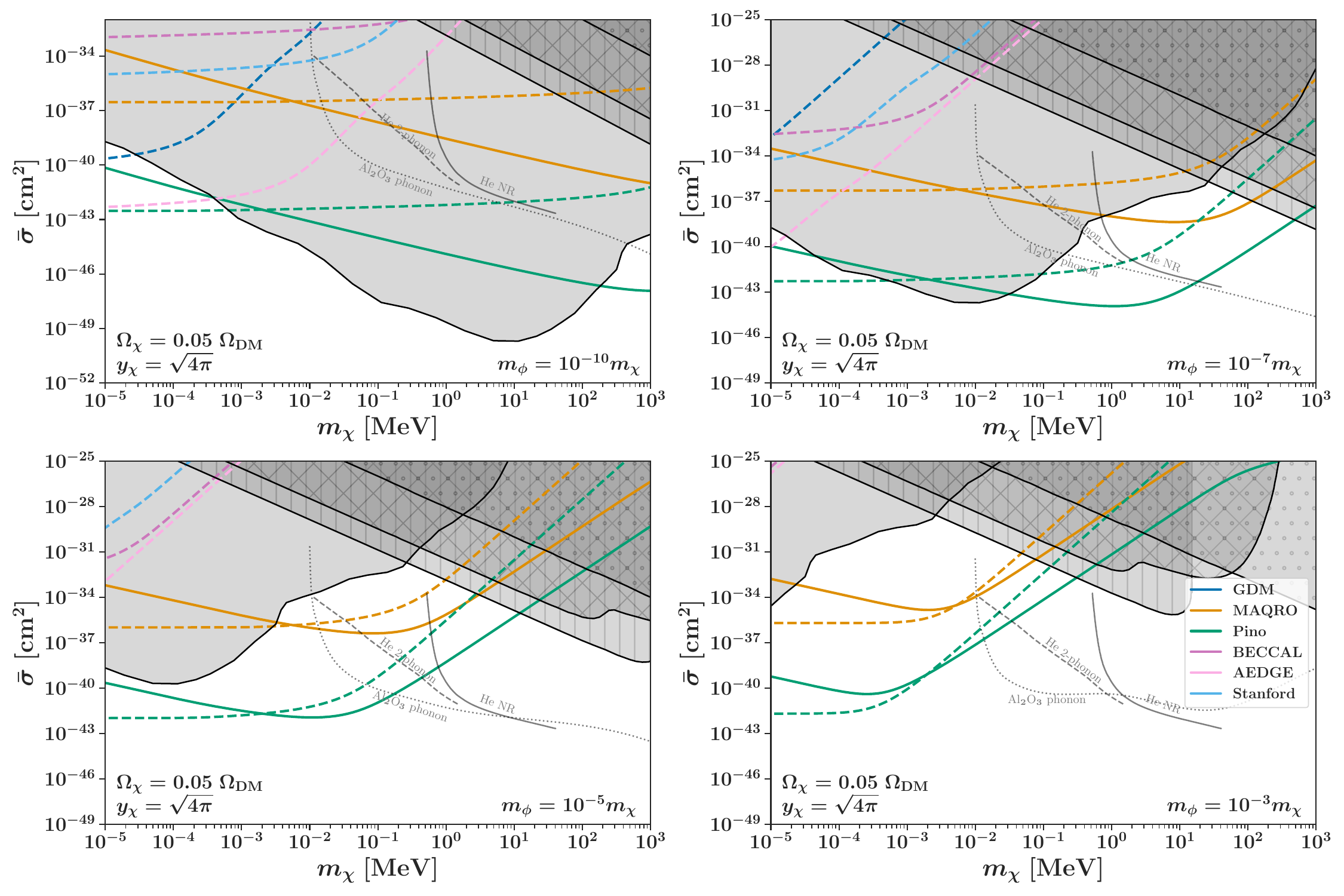}
    \caption{DM detection prospects for atom interferometers in the nucleon cross section vs $m_{\chi}$ plane assuming a light mediator, $m_\phi \propto m_\chi$. The solid, colored lines show the limits from decoherence, which apply to matter interferometers, and the dashed, colored lines show the limits from phase shifts, applying to both matter and cold atom interferometers. The different colors signify different experiments, where we assume 1 year of integration time for each experiment. The gray lines give the superfluid helium prospects for the nuclear recoil (solid) and 2-phonon (dashed) cases \citep{Knapen2017b}, and the Al$_2$O$_3$ prospects for phonon excitations (dotted) \citep{Griffin2020}. The superfluid helium forecasts assume a massless mediator. We assume an energy threshold of $\omega_{\rm{min}} = 1~\rm{meV}$ for Al$_2$O$_3$. The solid, gray regions indicate 5th force constraints from Ref.~\citep{Murata2015} (no hash), and stellar emission constraints from RGB stars (vertical hash)~\cite{Hardy:2016kme}, HB stars ('X' hash)~\cite{Hardy:2016kme}, and SN1987a (dot hash)~\cite{Ishizuka:1989ts}.} Although the atom interferometer experiments can probe to much lighter $m_\chi$ than shown here, we only plot to $10~\rm{eV}$. Lighter DM would have velocities greater than the escape speed of the galaxy, if we assume that the DM decouples before the QCD temperature. For a longer discussion of these other constraints, see Section~\ref{sec:other_constraints}. We assume that the DM particle $\chi$ is a DM subcomponent and set $y_\chi = \sqrt{4\pi}$ for the 5th force and stellar emission constraints.
    \label{fig:results}
\end{figure*}

\begin{figure*}[ht]
    \centering
    \includegraphics[width=\textwidth]{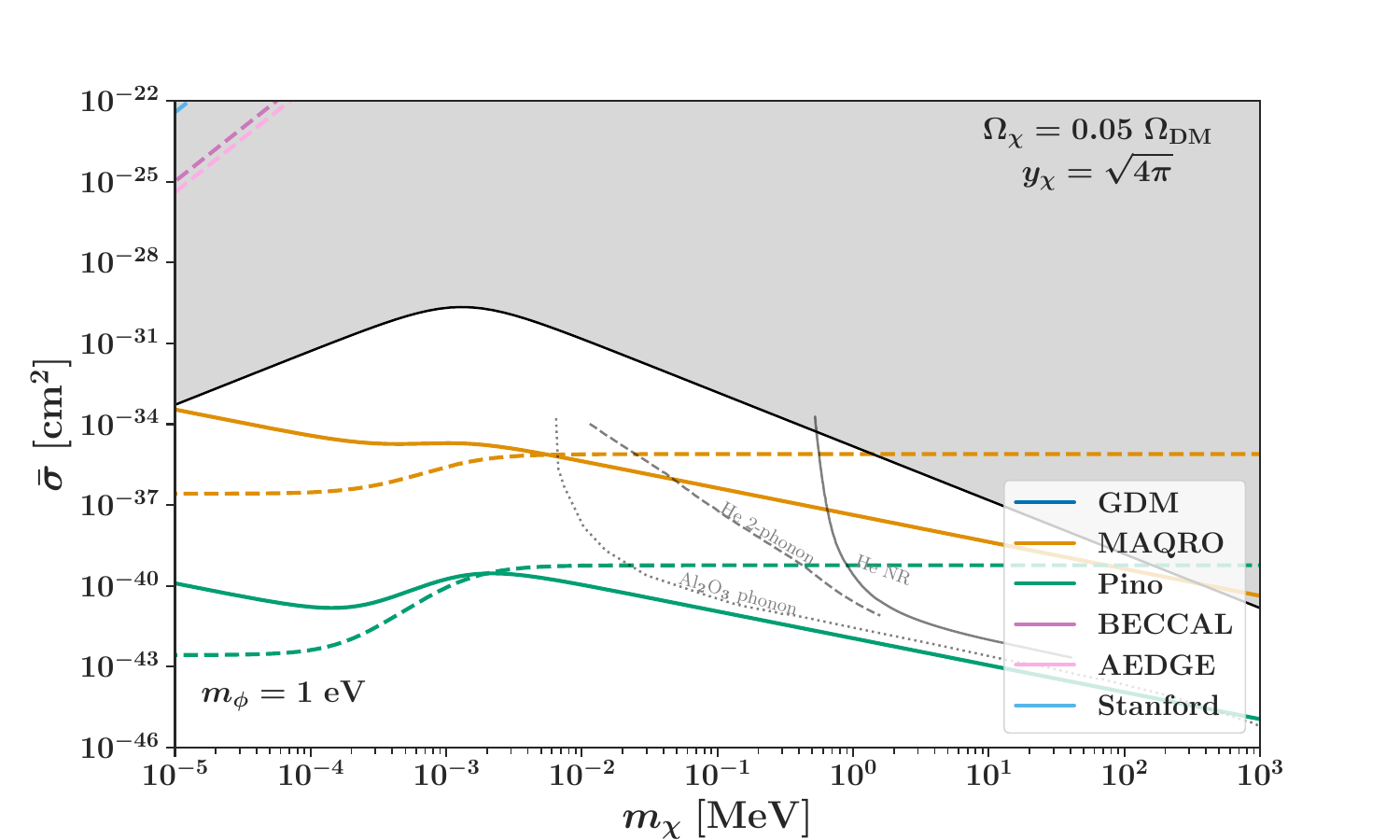}
    \caption{DM detection prospects for atom interferometers in the nucleon cross section vs $m_{\chi}$ plane for fixed $m_{\phi} = 1~\rm{eV}$. The solid, colored lines show the limits from decoherence, and the dashed, colored lines show the limits from phase shifts. The different colors signify different experiments, where we assume 1 year of integration time for each experiment. The gray lines give the superfluid helium prospects for the nuclear recoil (solid) and 2-phonon (dashed) cases \citep{Knapen2017b}, and the Al$_2$O$_3$ prospects for phonon excitations (dotted) \citep{Griffin2020}. The superfluid helium forecasts assume a massless mediator. We assume an energy threshold of $\omega_{\rm{min}} = 1~\rm{meV}$ for Al$_2$O$_3$. The solid, gray region indicates 5th force constraints from Ref.~\citep{Murata2015}. The change in the shape of the region is set by the transition from heavy to light mediator (as $m_\chi$ increases relative to the fixed $m_\phi = 1~\rm{eV}$).}
    \label{fig:results_fixedmphi}
\end{figure*}

\begin{figure}[ht]
    \centering
    \includegraphics[width=\textwidth]{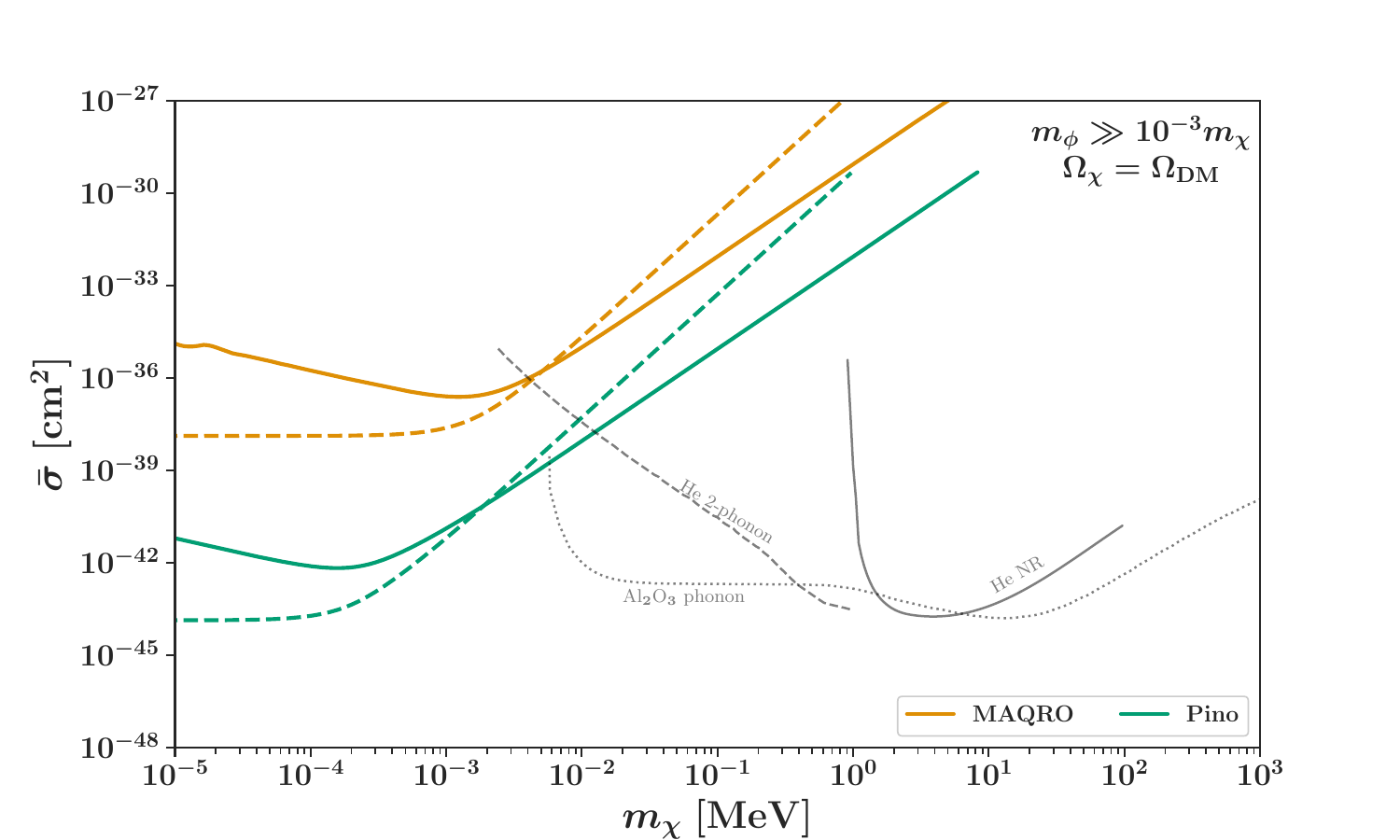}
    \caption{Forecasted DM-nucleon cross section limits from atom interferometers for the heavy mediator case.  The solid, colored lines show the limits from decoherence, and the dashed, colored lines show the limits from phase shifts. The different colors signify different experiments, where we assume 1 year of integration time for each experiment. The gray lines give the superfluid helium prospects for the nuclear recoil (solid) and 2-phonon (dashed) \citep{Knapen2017b}. The dotted line gives the Al$_2$O$_3$ prospects from phonon excitations \citep{Griffin2020}. We note that much of this parameter space is excluded by various collider, astrophysical, and cosmological constraints, dependent on the mediator mass and hence not shown here \citep[see Fig. 6 of Ref.][and related discussions therein]{Knapen2017b}.  For this plot, we also assume that the $\chi$ makes up all of the DM. Although the dark matter model space probed by atom interferometers for massive mediators is of somewhat limited interest due to existing collider, astrophysical, and cosmological constraints, we include this plot because it enables an easy calculation of the constraints for the other processes we consider in Section~\ref{sec:other_process}, where such constraints do not necessarily apply.}
    \label{fig:heavy_reach}
\end{figure}

Integrating over the imaginary part in this same way would give us zero. Some anisotropy in the DM flux is needed for its interaction to induce a relative phase between the two clouds. Taking into account that anisotropy, and anticipating the full derivation of the rate of DM interactions, we parameterize the effect responsible for generating a non-zero shift in the relative phase as:
\begin{equation}\label{eqn:phase_effect}
\begin{split}
  \frac{1}{2}\int_{-1}^{1} \mathrm{Im}(p_{\rm{decoh}})\,  e^{- \displaystyle \tilde v_e q  \Delta x \cos \theta_{\bf q \Delta x}}  ~d\cos\theta_{\mathbf{q\Delta x}} & = \frac{\tilde v_e \sin (q\Delta x) \cosh (\tilde v_e q \Delta x) - \cos (q \Delta x) \sinh(\tilde v_e q \Delta x)}{q\Delta x (\tilde v_e^2 +1)} \; ,
  \end{split}
  \end{equation}
where $\tilde v_e =  v_e / (v_0^2 m_\chi \Delta x)$ is a scaled version of the Earth's velocity, and we have assumed that ${\bf \Delta x}$ is aligned with ${\bf v_e}$. Like the decoherence effect, the phase effect is maximal at $q \sim 1/\Delta x$. However, it oscillates rapidly about $0$ for $q > 1/\Delta x$. Note that for small $q$, Eqn.~\ref{eqn:phase_effect} goes as $(q\Delta x)^2\tilde{v}_e$ (see the discussion in Section.~\ref{sec:nuc_recoils} for how this affects our constraints). Although this form factor vanishes when $q \to 0$, as the limits in Fig.~\ref{fig:results},~\ref{fig:results_fixedmphi} and~\ref{fig:heavy_reach} show, the sensitivity to the phase approaches a constant because the $q$ dependence cancels in the rate.

In this paper we only consider observables that get coherent enhancements when dark matter does not resolve the size of the atom interferometer~\cite{Badurina:2024nge}.
In these cases, the decoherence factor probability factorizes from the overall probability that an incoming scattering event produces a given final state. In other words, the dynamic structure factor can be written as:
\begin{equation}
    S(\mathbf{q}, \omega) \equiv \frac{p_{\rm{decoh}}}{V} \sum_f |\langle f | \mathcal{F}_T(\mathbf{q}) | i \rangle |^2~2\pi\delta(E_f - E_i - \omega_{\mathbf{q}}) \; ,
\end{equation}
where $\mathcal{F}_T$ is the target form factor and the delta function enforces energy conservation.

The last factor in determining the general rate equation is the DM phase-space distribution. We will follow Ref.~\citep{Trickle2020} and take:
\begin{equation} \label{eq:fv}
    f(\mathbf{v}) = \frac{1}{N_0} \exp\left(-\frac{(\mathbf{v} + \mathbf{v}_e)^2}{v_0^2}\right) \Theta(v_{\rm{esc}} - \|\mathbf{v}+\mathbf{v}_e\|) \; ,
\end{equation}
where $\|\mathbf{v}_e\| = 240~\rm{km/s}$ is the Earth's velocity\footnote{Technically, $\mathbf{v}_e$ should be replaced by the velocity of each experiment, for the space-based experiments. However, many of these mission concepts do not have finalized orbits. They are likely to be solar orbits with semi-major axis $\sim 1~\rm{AU}$. To first order in orbital eccentricity, this would give them similar velocities to the Earth. The experiment on the International Space Station (BECCAL) is one exception to this. However, it's orbital velocity around the Earth is within $\lesssim 10\%$ of the Earth's velocity we assume here. Thus we assume Earth's velocity for all of our experiments.}, $\Theta(x)$ is the Heaviside function, $v_{\rm{esc}} = 600~\rm{km/s}$ is the escape velocity of the Galaxy, and $N_0$ is the normalization factor:
\begin{equation}
    N_0 = \pi^{3/2} v_0^3 \left[\mathrm{erf} \left( \frac{v_{\rm{esc}}}{v_0}\right) - \frac{2}{\sqrt{\pi}} \frac{v_{\rm{esc}}}{v_0} \exp\left(- \frac{v_{\rm{esc}}^2}{v_0^2} \right) \right] \; .
\end{equation}

We can simplify the integration by remembering that there is a relationship between the velocity and the momentum transfer. Regardless of the target, the energy deposition from a given DM momentum transfer and velocity is:
\begin{equation}\label{eqn:nr_energy_dep}
    \omega_{\mathbf{q}} = qv\cos\theta_{\mathbf{qv}} - \frac{q^2}{2m_{\chi}} \; ,
\end{equation}
where $\theta_{\mathbf{vq}}$ is the angle between $\mathbf{v}$ and $\mathbf{q}$, and $v = \|\mathbf{v}\|$.

Thus, the energy deposition delta function can be traded for one in velocity. In general, we can define \cite{Trickle2020}:
\begin{align}
\begin{split}
    g(\mathbf{q}, \omega) &\equiv \int d^3\mathbf{v}~f(\mathbf{v}) 2\pi \delta(\omega-\omega_{\mathbf{q}}) \\
    &=
    \frac{2 \pi^2 v_0^2}{N_0 q} \left[\exp \Big( - v_-^2 (\mathbf{q}, \omega)  /v_0^2 \Big) - \exp (- v_{\rm esc}^2/v_0^2)\right] \; ,
\end{split}
\label{eq:gvel}
\end{align}
where:
\begin{equation}
    v_-(\mathbf{q},\omega) = \mathrm{min} \left[ \frac{1}{q}\left | \mathbf{q}\cdot\mathbf{v}_e + \omega + \frac{q^2}{2m_{\chi}} \right|, v_{\rm{esc}} \right] \; .
    \label{eq:vmin0}
\end{equation}
The energy deposition in this equation is set by the conservation of energy equation in the dynamic structure factor, which is different for the various channels we discuss in this paper.

The rate equation for DM scattering that is measurable via either decoherence or phase shifts is then:
\begin{equation}\label{eqn:rate}
    R = \frac{1}{\rho_TV}\frac{\rho_\chi}{m_\chi} \frac{\pi \bar{\sigma}}{m_{\chi}^2}\int \frac{d^3\mathbf{q}}{(2\pi)^3}\mathcal{F}_{\rm{med}}^2({\bf q}) \left(1 - \exp(i\mathbf{q}\cdot\mathbf{\Delta x})\right)\sum_f |\langle f | \mathcal{F}_T(\mathbf{q}) | i \rangle |^2 g(\mathbf{q}, E_f - E_i) \; .
\end{equation}

To get the total decoherence factor, we integrate over the measurement time and multiply by the total target mass. If we ignore any daily modulation, then there is no explicit time-dependence in the rate. This then gives a total decoherence of:
\begin{equation}\label{eqn:overall_dec}
    s-i\phi = \frac{1}{N_{\rm ind}} \frac{\rho_\chi}{m_\chi} \frac{\pi \bar{\sigma}}{m_{\chi}^2} t_{\rm{exp}}\int \frac{d^3\mathbf{q}}{(2\pi)^3} \mathcal{F}_{\rm{med}}^2({\bf q})  \left(1 - \exp(i\mathbf{q}\cdot\mathbf{\Delta x})\right) \sum_f |\langle f | \mathcal{F}_T(\mathbf{q}) | i \rangle |^2 g(\mathbf{q}, E_f - E_i)\; ,
\end{equation}
where $t_{\rm{exp}}$ is the measurement time, and the volume factor in the rate equation cancels with the volume factor in the dynamic structure factor.
The real part of the above equation quantifies the decoherence induced in a  matter interferometer, while the imaginary part applies to both matter interferometers and diffuse clouds.

We stress here that, unlike other direct detection experiments, \textbf{there is no minimum energy threshold for these atom interferometer experiments}. Unlike traditional direct detection setups, atom interferometers do not directly measure the energy deposition -- only the visibility and phase. This allows atom interferometers to probe much lighter DM masses than traditional direct detection setups. In practice, the decoherence observable loses sensitivity to light DM once the mass is below scales $m_\chi v_0 \lesssim 1/\Delta x$. On the other hand, the phase has no minimum. The main limitations for this method comes from other backgrounds.

There are many possible decoherence and phase shift mechanisms for atom interferometers \citep[e.g., see Ref.][]{LeGouet2008}. To definitively make a DM detection, an atom interferometer would need to see a signal that varies with the expected DM flux \citep{Riedel2017}. For a terrestrial experiment, this would give the so-called ``daily modulation" caused by the Earth's rotation with respect to the incoming DM flux. For a space-based experiment, like most of the ones we consider here, the exact modulation will depend on the orbit. Low-Earth orbits will lead to modulation signals with timescales on the order of the orbit (\textit{e.g.,} for an experiment on the International Space Station, the timescale is $\sim{90}$ minutes). Other orbits would need to be studied in depth. We do not consider this effect in-depth -- we assume that if the decoherence and phase effects can be measured in a single shot, then their variance with time will also be measurable (but see Appendix~\ref{sec:appendix2} for a short description of how to include the daily modulation). Thus, we study here a necessary, but not sufficient, metric for detecting DM with atom interferometers.

We say that an atom interferometer has sensitivity to a given DM model when the estimated signal is larger than the expected noise (\textit{i.e.,} we set a signal-to-noise threshold of 1). Thus, the overall sensitivity to a given cross section is given by solving for the cross section in the equation:
\begin{equation}\label{eqn:overall_sens}
    \frac{(X - X_{\rm{bkg}})^2}{\sigma_X^2} = 1 \; ,
\end{equation}
where $X$ denotes either the visibility, $V$, or the phase, $\phi$. $X_{\rm{bkg}}$ is the average value of either observable without any DM scattering effects, while $\sigma_X$ is the noise for each observable. We note that we do not study backgrounds here such as cosmic rays and solar photons. %While these events will produce decoherence, we expect that the rates will be highly suppressed since the higher momentum transfer limits their ability to benefit from the Born enhancement effect.  One may be concerned that the tail of the interaction rate at very low momentum transfer could be a significant background to a dark matter search;
Ref.~\cite{Kunjummen2022} claimed, via an order of magnitude estimate, that the decoherence between atom clouds caused by cosmic rays are not a significant background.  Solar photons, in the forward scattering limit, could be a background source for phase shifts due to their lower energies (which can mimic dark matter kinematics) and high fluxes.  However, their interaction rate will be suppressed because the atom clouds are neutral, with interactions occurring through the polarizability, which is generally suppressed.  We leave detailed study for future work, but note in the interim that our reach curves can be simply re-scaled with the background event rate.

First, we consider the decoherence sensitivity. In the absence of decoherence, environmental effects, and systematics, $V = 1$. The sensitivity to any visibility change, per shot, is limited by the irreducible quantum noise limit (QNL),
\begin{equation}\label{eqn:phi_sens_N}
    \sigma_V \equiv \frac{1}{\sqrt{N_{\rm ind}}} \; .
\end{equation}
Modern atom interferometers have demonstrated detection sensitivity on the QNL level (see, \textit{e.g.},~\cite{bize2005cold}). In this work, we assume that all proposed missions could operate at the quantum limit, and thus the noise of the visibility is given by the equation above. We also assume that $\sigma_V$ will scale with the number of measurements, $N_{\rm{meas}}$, as: $\sigma_V \propto N_{\rm{meas}}^{-1/2}$.  As we discuss further in the next subsection, $N_{\rm meas}$ is the number of drop times accumulated over the run-time (assumed to be 1 year) of the experiment.

% \sout{ If there is no decoherence, then $V = 1$; however, most interferometers have at least some decoherence from environmental effects. A conservative fringe visibility factor is $V_\text{bkg} = 0.5$ \citep{Riedel2017}, which we will use for all of our experiments. We assume that, per measurement, the visibility noise is $\sigma_V/V = 0.1$.}

Now let us consider the phase sensitivity. We ignore any effects from gravity here and set the average $\phi_{\rm{bkg}} = 0$. The standard deviation of the phase is then the minimum measurable phase shift, which is set by the number of nucleons, typical contrast, and other backgrounds. The experiments we consider do not directly list their minimum phase shifts. However, in the case of zero systematic issues, the phase sensitivity can be estimated by QNL as well \citep{Itano1993, Sorrentino2014}. We define our phase error as:
\begin{equation}\label{eqn:phi_sens_N}
    \sigma_\phi \equiv \frac{1~\rm{rad}}{\sqrt{N_{\rm ind}}} \; .
\end{equation}
Note that we assume perfect contrast when setting the phase constraints. We list the minimum measurable phase shifts for each mission in the last column of Table~\ref{tab:mission_params}. As in the case of the visibility, the phase sensitivity will also scale with $N_{\rm{meas}}$ as: $\sigma_\phi \propto N_{\rm{meas}}^{-1/2}$.

We note that the minimum measurable acceleration can also be used to calculate the minimum phase via:
\begin{equation}\label{eqn:phi_sens}
    \sigma_\phi = \frac{1}{4}m_{\rm ind} \Delta x \, t_\text{exp} a_\text{min},
\end{equation}
where $m_{\rm ind}$ is the mass of one independent particle within the atom interferometer, and the momentum injected by the laser beam has been written as a function of the separation of the wave packets and the measurement time. The factor of $1/4$ accounts for our use of the full measurement time, $t_{\rm{exp}}$, rather than the time between pulses ($\frac{1}{2}t_{\rm{exp}}$), which is typically used in these estimates.

We would also like to note: there are \textit{upper} bounds on the sensitivities as well: if there is too much decoherence from DM scattering in an experiment, then the experiment would never measure a fringe! This is then degenerate with an experiment simply not working properly (\textit{i.e.}, not properly producing a coherent state or having too many decoherence-producing backgrounds). In principle, the space-based experiments are more subject to this problem than terrestrial ones - if DM scattering cross sections were high enough to always produce decoherence, then they would most likely never reach the terrestrial detectors due to atmospheric shielding. Thus, a null test for DM in this case would be to compare identical setups in Earth and space. We ignore these issues in our sensitivity curves since this would be easily solved via the above test. In addition, several atom interferometers have been shown to properly work on the surface of the Earth, and one has been shown to work in Low-Earth Orbit \citep{Aveline2020, Lachmann2021}. Although it may be difficult to distinguish DM from other backgrounds, no signal of decoherence unambiguously constrains DM interactions.

To summarize: DM scattering with momentum transfer along the separation axis of an atom interferometer causes decoherence and phase shifts in an atom interferometer. The rate of these interactions is given by Eqn.~\ref{eqn:rate}. The exact form of the rate will depend on the particulars of the DM model and scattering type. The measured decoherence and phase shift effects are then given by Eqn.~\ref{eqn:overall_dec} for the cases we consider. The overall sensitivity of the experiment is given by comparing either the decoherence or the phase shift to any background signal and the expected noise via Eqn.~\ref{eqn:overall_sens}. The resulting limits are shown in Figs.~\ref{fig:results}, \ref{fig:results_fixedmphi}, and \ref{fig:heavy_reach}, which we will discuss in detail in Sections~\ref{sec:nuc_recoils} \& \ref{sec:other_process}.

\subsection{Atom Interferometer Experiments}\label{sec:exps}

In this section, we describe the various atom interferometer experiments that we consider in this paper. For the most part, these are future, space-based experiment concepts that span a wide range in the parameters that matter for our DM signals ($\Delta x$, $N_{\rm{nucleon}}$, $r_{\rm{cloud}}$, and $t_{\rm{exp}}$). Many of the other experiments that we do not discuss here (\textit{e.g.,} MAGIS-100~\citep{MAGIS-100:2021etm}, AION \cite{Badurina2020}, STE-QUEST \cite{Aguilera2014}) have parameters, where published, within the ranges we consider here and also feature multiple interferometers which are beyond the scope of the present work. We note that our public code can be used to calculate the limits for any other experiment, given the necessary parameters.

Table~\ref{tab:mission_params} gives the important parameters for all of the experiments we consider and we discuss each of them in turn below. In analogy to other direct detection experiments, which typically give limits assuming 1 kg-yr of exposure, we assume an overall integration
time of 1 year for all experiments. To calculate the number of measurements that each experiment will take, $N_{\rm meas}$, we assume no downtime between measurements and simply divide the assumed 1 year
integration time by the experimental measurement time.

The Bose-Einstein Condensate and Cold Atom Laboratory (BECCAL) is a proposed upgrade \citep{Frye2021} to the Cold Atom Lab (CAL), which is currently running on the International Space Station (ISS) \citep{Elliott2018, Aveline2020}. This upgrade would improve the current atom interferometer capabilities of CAL, including a higher number of atoms in the condensates. BECCAL would study a few different atoms; however for simplicity we focus on its $^{87}$Rb capabilities in this paper. We assume that BECCAL would produce condensates with $10^6$ atoms, a cloud size of $150~\mu\rm{m}$, and a free-fall (measurement) time of 2.6 s. We take the separation between the atom interferometer arms to be $\Delta x = 3~\rm{mm}$. Given the number of atoms in this experiment, we calculate a sensitivity of: $\sigma_\phi = 10^{-3}$.

Macroscopic Quantum Resonators (MAQRO) is a proposed space mission to perform high-mass matter interferometry \citep{Kaltenbaek2015, Kaltenbaek2016}. The mission would use SiO$_2$ molecules in a cloud with $10^{10}$ nucleons and a radius of $120~\rm{nm}$. The baseline separation would be $100~\rm{nm}$ and the free-fall measurement time would be $100~\rm{s}$. Note that since this experiment uses a solid nanoparticle rather than a diffuse atomic cloud, there is only $1$ independent phase measurement per drop. Thus, we assume a phase sensitivity for this experiment of: $\sigma_\phi = 1$.

The Gravity Probe and Dark energy Detection mission (GDM) is a NASA Innovative Advanced Concepts (NIAC) Phase II mission concept \cite{GDM_NIAC, GDM}. Although its primary goal would be to measure deviations from the gravitational inverse-square law, its instruments would be well-suited for the direct detection scheme we consider here. It would consist of a constellation of four spacecraft, each with six atom interferometers. In this paper, we consider the limits from just one of these interferometers. However, the large number of interferometers, each with separation axes in different directions would be useful for disentangling any possible DM signature from other backgrounds. This is a futuristic mission concept and many of the parameters are not concretely set yet. We take the same values assumed by the GDM team in their benchmark forecasts \citep{Shengwey}. We assume an atom interferometer baseline of $\Delta x = 25~\rm{m}$, an interrogation time of $20~\rm{s}$, a cloud size of $r_{\rm{cloud}} = 1~\rm{mm}$, and $10^8$ $^{87}$Rb atoms. From the number of atoms, we assume a phase sensitivity of $\sigma_\phi = 10^{-4}$.

Another mission is the Atomic Experiment for Dark Matter and Gravity Exploration in Space (AEDGE)~\citep{AEDGE:2019nxb}, a space-based version of the long baseline earth proposals MAGIS-100~\citep{MAGIS-100:2021etm} and AION~\citep{Badurina2020}. According to the quoted phase sensitivity of $10^{-5} \text{rad} \, \text{Hz}^{-1/2}$, AEDGE will consist of a diffuse cloud of $10^{10}$ $^{87}$Sr atoms with a radius of $3$ mm. The largest separation between the two clouds is expected to be 0.9 m, each shot lasting 10 minutes. We expect a phase sensitivity of $\sigma_\phi = 10^{-5}$.

We also consider a proposed terrestrial experiment, described in Ref.~\citep{Pino2018}, which we refer to as the ``Pino" experiment in this paper. This table-top experiment concept would use an all-magnetic scheme to perform a double slit experiment with a macroscopic sphere. We take the example experimental parameters listed in Appendix~A of Ref.~\citep{Pino2018} as our parameters. Namely, we assume a sphere of Niobium with radius 1 micron and $2\e{13}$ nucleons. We assume a separation of $290~\rm{nm}$ and a free-fall time of $0.483~\rm{s}$. The goal for this experimental concept is to measure the decoherence from the self-gravity of the sphere, but it is not clear if this experiment could produce a phase measurement as well. Here we assume it will be able to measure a phase, but like MAQRO, the number of independent phase measurements per shot is $N_{\rm{ind}} = 1$. Then its phase sensitivity is: $\sigma_\phi = 1$.

Finally, we consider a current terrestrial experiment, an atom interferometer drop tower described in Refs.~\citep{Asenbaum2020, Overstreet2022}. This experiment, which we refer to as the ``Stanford" experiment in this paper, drops $^{87}$Rb atoms in a 10~m tower to perform the interferometry measurements. They use $4\e{6}$ atoms and have a free-fall time of $1.91~\rm{s}$. The cloud radius is $2\e{-4}~\rm{m}$ and the separation between the clouds is $0.067~\rm{m}$. From the number of atoms, we assume a phase sensitivity of $\sigma_\phi = 5.0\e{-4}$. This is within an order of magnitude of the phase calculated via Eqn.~\ref{eqn:phi_sens} and their reported acceleration sensitivity per shot of $a_{\mathrm{min}} = 1.4\e{-10}~\rm{m/s}^2$.

\begin{table*}[hb]
    \centering
	\begin{tabular}{|l|c|c|c|c|c|c|c|}
        \hline
        Mission & Type & Target & $r_{\rm{cloud}}$ & $N_{\rm{nucleon}}$ & $\Delta x$ & $t_{\rm{exp}}$ & $\sigma_\phi$\\
        & & & [m] & & [m] & [s] & [rad]\\
        \hline
        \hline
        MAQRO \citep{Kaltenbaek2015, Kaltenbaek2016} & Solid &SiO$_2$ & $1.2\e{-7}$ & $10^{10}$ & $10^{-7}$ & $100$ & $1.0$ \\
        Pino \citep{Pino2018} & Solid & Nb & $10^{-6}$ & $2.2\e{13}$ & $2.9\e{-7}$ & 0.483 & $1.0$ \\
        BECCAL \citep{Frye2021, Elliott2018, Aveline2020} & BEC & $^{87}$Rb & $1.5\e{-4}$ & $8.7\e{7}$ & $3\e{-3}$ & 2.6 & $1.0 \times 10^{-3}$ \\
        GDM \citep{GDM, GDM_NIAC, Shengwey} & BEC & $^{87}$Rb & $10^{-3}$ & $8.7\e{9}$ & $25$ & $20$ & $1.0 \times 10^{-4}$ \\
        Stanford \citep{Asenbaum2020, Overstreet2022} & Diffuse cloud & $^{87}$Rb & $2\times 10^{-4}$ & $3.5\e8$ & $6.7\e{-2}$ & $1.91$ & $5.0 \times 10^{-4}$\\
        AEDGE \citep{El-Neaj2020} & Diffuse cloud& $^{87}$Sr & $3\times 10^{-3}$ & $8.7\e{11}$ & $0.9$ & $600$ & $1.0 \times 10^{-5}$\\
        \hline
    \end{tabular}
    \caption{Mission concept parameters assumed in this paper. The phase sensitivity, $\sigma_{\phi}$, is given per exposure time. We calculate $\sigma_{\phi}$ for all experiments based on Eqn.~\ref{eqn:phi_sens_N}. The other parameters are taken from the references shown next to each experiment. The details are given in Section~\ref{sec:exps}.}
    \label{tab:mission_params}
\end{table*}

\section{Dark Matter Interaction Rates in Atom Interferometers}\label{sec:dm_rates}

Various interaction processes between the DM and atoms can cause the decoherence and phase shift effects. As will be discussed in detail below, because atoms in an atom interferometer begin in a coherent state, the DM-atom interaction rate will receive an $N^2$ enhancement in the coherent scattering limit in the scenarios we consider (contrast loss for matter interferometers, phase-shifts for matter interferometers and cold atom clouds), where $N$ is the number of nucleons or electrons in the target. Thus, to exploit this $N^2$ enhancement, we primarily seek processes that do not break the coherence of states within one wavepacket. In particular, we consider DM scattering via nuclear recoils (\ref{sec:nuc_recoils}), and interactions involving dark photons and coherent axion scattering (\ref{sec:other_process}). Only the nuclear recoils produce interesting constraints; however, we give estimates of the constraints for the other cases.

\subsection{Nuclear Recoil}\label{sec:nuc_recoils}

In this section, we derive the atom interferometer observables expected from DM scattering via nuclear recoils. We must use the physics of nuclear recoils to set two of the terms in Eqn.~\ref{eqn:overall_dec}: the target form factor, and the energy deposition via the conservation of energy. We start with these definitions and then walk through the simplifications needed to arrive at the final decoherence equation, Eqn.~\ref{eqn:nr_dec}. We then consider the phase observable.

We will be agnostic about the mediator in this section and take its form factor to be:
\begin{equation}\label{eqn:med}
    \mathcal{F}_{\rm{med}}(q) = \frac{(m_\chi v_0)^2+m_\phi^2}{q^2 + m_\phi^2} \; ,
\end{equation}
where $v_0$ is the DM velocity dispersion. This reduces to the light mediator form factor ($\mathcal{F}_{\rm{med}}(q) \rightarrow (m_\chi v_0/q)^2$) for $m_{\phi} \rightarrow 0$, and to the heavy mediator form factor ($\mathcal{F}_{\rm{med}}(q) \rightarrow 1$) for $m_{\phi} \gg q$.

An important factor in setting these limits is whether the scattering is in the coherent limit (\textit{i.e.}, whether the DM scatters off of the cloud as a whole or only 1 nucleus). This is set by the momentum transfer, $q$, of the dark matter and the relevant cloud radii. We note that the various limits here will apply, when considering loss of contrast, to  matter interferometers, and when considering phase-shifts, to both matter and cold atom interferometers. These limits will be important for both finding the target form factor and the minimum velocity, $v_-$.

Consider an atom interferometer that has $N_A$ identical atoms, each with atomic mass number $A$. The structure factor contains the response of these $N=AN_A$ nucleons to a DM scattering event. It will depend on the momentum transfer, $q$, versus the size of the atom cloud, $r_C$, the size of each atom, $r_A$, and the size of each nucleon, $r_N$. In particular, there are four regimes to consider:
\begin{enumerate}
    \item $q \ll 1/r_C$: Here, the DM does not ``see'' any of the separate atoms. Instead, it sees the entire atomic cloud as one object. This means that the rate will receive a Born enhancement: $R \propto N^2$.
    \item $1/r_C < q < 1/r_A$: In this case, the DM sees the separate atoms, but does not see individual nucleons. Thus, there is a Born enhancement for each of the nucleons within an atom, but not for all of the atoms: $R \propto A^2 N_A$.
    \item $1/r_A < q < 1/r_N$: Here, the DM sees each separate nucleon. There is no Born enhancement of any kind: $R \propto N$.
    \item $q \gg 1/r_N$: We do not consider any extra interactions in this case, and we allow the rate to go to zero whenever the momenta is large enough to probe within individual nucleons.
\end{enumerate}

With all of this in mind, we take the structure factor to be:
\begin{equation} \label{eq:S}
    \sum_f |\langle f | \mathcal{F}_T(\mathbf{q}) | i \rangle |^2 = \sum_{i,j=1,...,N_A} \langle e^{-i\Delta \mathbf{y}_{ij}\cdot \mathbf{q}} \rangle_{\rm{target}} = N[1 + AF_A^2(q) + A(N_A - 1) F^2(qr_C)] \; ,
\end{equation}
where $F_A(x)$ is the atomic form factor, $F(x)$ is the cloud form factor, and $\Delta y_{ij}$ are the interparticle spacings. We take the atomic form factor to be the Helm form factor \citep{Coskuner2019}:
\begin{equation}
    F_A(q) = \frac{3j_1(qr_A)}{qr_A}e^{-q^2s_p^2/2} \; ,
\end{equation}
where $j_1(x)$ is a spherical Bessel function, $r_A\approx A^{1/3} \times 1.2~\rm{fm}$ is the typical size of an atom, and $s_p \approx 0.9~\rm{fm}$ is the skin depth.

The cloud form factor is typically taken as \citep{Riedel2013, Afek2021}:
\begin{equation}
    F(x) = \frac{3j_1(x)}{x} \; .
\end{equation}
Note that for low momentum transfer, $F^2(q \, r_C) \rightarrow 1$. For high momentum transfer, $F^2(q \, r_C) \rightarrow 1/(q \, r_C)^{4}$. This cloud form factor is appropriate for experiments with macroscopic particles, such as Pino and MAQRO, as well as diffuse atomic clouds, such as Stanford and AEDGE. However, for the BEC experiments we discuss in this paper (BECCAL, and GDM), although we treat them exactly in the same way as a diffuse cloud under dark matter scattering, we must make a small modification. As discussed in Appendix~\ref{sec:bec_appendix}, the form factor for the BEC experiments is: $F(x) = \exp\left[-(qr_C/2)^2\right]$. Note that although the above form factor is a reasonable approximation to this form factor, we do use this BEC form factor for the BEC experiments.

Now, let us consider the conservation of energy. For a given momentum transfer $q$, the change in energy of a target nucleus is: $E_f - E_i = q^2/(2m_i)$, where $i$ refers to either the full atomic cloud or one atom, as discussed above. However, note that for the dark matter masses that we consider in this paper $m_\chi \ll m_i$ regardless of the target $i$. Thus, we can approximate $E_f - E_i = \omega \simeq 0$. With the energy deposition given by Eqn.~\ref{eqn:nr_energy_dep}, the minimum velocity is then given by:
\begin{equation}\label{eqn:vmin}
    v_{-}({\bf q}) \approx \mathrm{min} \left[ \frac{1}{q}\left | \mathbf{q}\cdot\mathbf{v}_e + \frac{q^2}{2m_{\chi}} \right|, v_{\rm{esc}} \right]\; .
\end{equation}

Note that $v_{-}$ depends on both the magnitude and direction of $\mathbf{q}$. Although the direction will not have a large impact on the decoherence rate calculation, it will be important for the phase calculation.

\subsubsection{Decoherence}
The limits of this subsection apply solely to matter interferometers. The decoherence is given by the real part of Eqn.~\ref{eqn:overall_dec}.  As was shown by Ref.~\citep{Riedel2017}, the daily modulation only affects the decoherence rate by a factor $2$. We also find that it has little effect on the final limits (see Appendix~\ref{sec:appendix2}). Thus, for ease of understanding the underlying physics, we set $\mathbf{v}_e = 0$ for the decoherence calculations.

Putting all of the above together gives a decoherence rate of:
\begin{align}
\begin{split}
    R_{\rm{decoh}} = \frac{1}{\rho_TV}\frac{\rho_\chi}{m_\chi} \frac{\pi \bar{\sigma}v_0^2(q_0^2+m_\phi^2)^2}{2N_0m_{\chi}^2m_{\phi}^4}\int dq \frac{q}{\left((q/m_{\phi})^2 + 1\right)^2} N[1 + AF_A^2(q) + A(N_A - 1)F^2(qr_C)]\\
    \times \left[\exp\left(-\frac{v_-^2(q)}{v_0^2}\right) - \exp\left(- \frac{v_{\rm esc}^2}{v_0^2}\right)\right] \int_{-1}^1 d\cos\theta \left(1-\cos(\mathbf{q}\cdot\mathbf{\Delta x})\right) \; ,
    \end{split}
\end{align}
where we define $\theta$ as the angle between ${\bf q}$ and ${\bf \Delta x}$.
The cosine integral can be done analytically, yielding:
\begin{align}\label{eqn:nr_rate}
\begin{split}
    R_{\rm{decoh}} = \frac{1}{\rho_TV}\frac{\rho_\chi}{m_\chi} \frac{\pi \bar{\sigma}v_0^2(q_0^2+m_\phi^2)^2}{N_0m_{\chi}^2m_{\phi}^4}\int dq \frac{q}{\left((q/m_{\phi})^2 + 1\right)^2} N[1 + AF_A^2(q) + A(N_A - 1)F^2(qr_C)]\\
    \times \left[\exp\left(-\frac{v_-^2(q)}{v_0^2}\right) - \exp\left(- \frac{v_{\rm esc}^2}{v_0^2}\right)\right] \left( 1 - \frac{\sin(q\Delta x)}{q\Delta x}\right).
\end{split}
\end{align}

The overall decoherence is given by plugging this into Eqn.~\ref{eqn:overall_dec}:
\begin{align}\label{eqn:nr_dec}
\begin{split}
    s = \frac{1}{N_{\rm ind}}  \frac{\rho_\chi}{m_\chi} \frac{\pi \bar{\sigma}v_0^2(q_0^2+m_\phi^2)^2}{N_0m_{\chi}^2m_{\phi}^4}t_{\rm{exp}}\int dq \frac{q}{\left((q/m_{\phi})^2 + 1\right)^2} N[1 + AF_A^2(q) + A(N_A - 1)F^2(qr_C)]\\
    \times \left[\exp\left(-\frac{v_-^2(q)}{v_0^2}\right) - \exp\left(- \frac{v_{\rm esc}^2}{v_0^2}\right)\right] \left( 1 - \frac{\sin(q\Delta x)}{q\Delta x}\right).
    \end{split}
\end{align}

To build intuition, we now consider various approximations of this integral. For each of these, we will also consider two scenarios for the mediator mass: $m_\phi \propto m_{\chi}$ and fixed $m_\phi$. These correspond to Figures~\ref{fig:results} \& \ref{fig:results_fixedmphi}, respectively.

\begin{enumerate}
    \item $m_{\chi} \rightarrow 0$ -- For either mediator case, this limit implies that both $q \, r_C \ll 1$ and $q \, \Delta x\ll 1$. The first of these allows the last term in the target form factor to dominate, which makes the rate go as $N^2$. The second of these will set the last term in Eqn~\ref{eqn:nr_rate} to go as $(q\Delta x)^2$ (to second order in the Taylor expansion). In addition, $v_- \rightarrow 0$. Taking the light mediator limit of the form factor, we find that the limit on the cross section scales as:
    \begin{equation}
        \lim_{m_\chi \to 0} \bar{\sigma}_n \propto \frac{\sqrt{N_{\rm ind}}}{N^2}\frac{1}{m_\chi (\Delta x)^2} ,
    \end{equation}
    where we only write the factors that depend on the DM mass and the experiment explicitly. This behavior applies to both $m_\phi \propto m_\chi$ and $m_\phi = \text{constant}$. In this limit of $m_\chi \to 0$, these two mediator scenarios correspond to the massless mediator and heavy mediator cases, respectively; however, both have the same form factor limit: $\lim_{m_\chi \to 0}{\cal F}_\text{med} \sim 1$.

 %   Note that the figures do not show this scaling for the GDM experiment. Because of its large $\Delta x$, there is a range of low $m_\chi$ where $q \Delta x \gtrsim 1$, and thus the DM will still cause decoherence. In this regime, the bounds from GDM then scale as $\bar \sigma \propto m_\chi \, \sqrt{N_{\rm ind}} / N^2$, as shown in the figures. At low enough $m_\chi$, below the scales shown in the figures, GDM would then follow the above scaling.

    \item $m_\chi \to \infty$ -- For $m_\phi \propto m_\chi$, the rate benefits from the momentum transfer growing proportionally to the dark matter mass. 
    In this limit, $q \, r_C \gg 1$ and $q \, \Delta x \gg 1$. The first of these means that the first term in the target form factor dominates, so $R_{\rm{decoh}}\propto N$. The second of these will set the probability of causing decoherence to $1$. Taking the light mediator limit of the form factor we find:
    \begin{equation}
        \lim_{m_\chi \to \infty} \bar{\sigma}_n \propto 
        \frac{\sqrt{N_{\rm ind}}}{N} m_{\chi}
        \quad \rm{for~}m_\phi \propto m_\chi \; .
    \end{equation}
    Thus the cross section limit at high masses is only dependent on the number of nuclei in the experiments and is largely insensitive to the rest of the experimental setup. Note that in Fig.~\ref{fig:results} we do not reach high enough masses to see this behavior for the experiments plotted. Instead, the higher masses in our plots in Fig.~\ref{fig:results} are explained by limit 3.

    Now let us consider the $m_\phi = \text{constant}$ case. We will set $m_\phi = 1 \text{ eV} \gtrsim 1/r_C$, as we do in Fig.~\ref{fig:results_fixedmphi}. Unlike in the previous case, for a fixed mediator mass the rate benefits from low momentum transfer, $q < r_C^{-1}$. Then, we expect the following scaling for the masses $m_\chi > m_\phi / v_0 \sim 1~\rm{keV}$:
    \begin{equation}
        \bar \sigma \propto 
        \frac{\sqrt{N_{\rm ind}}}{N^2}
        \frac{m_\phi^4 r_C^2}{ m_\chi} \quad \rm{for~}m_\phi = \text{constant} \; .
    \end{equation}
    Naively it might be surprising that the limits on the reference cross section improve linearly with $m_\chi$, as Fig.~\ref{fig:results_fixedmphi} displays. However, because we define $\bar \sigma \propto (y_\chi y_\phi)^2/ m_\chi^2$ for $m_\chi \gg m_\phi$ (see Eq.~\ref{eqn:RefXSec}), the constraints on the couplings $(y_\chi y_\phi)^2$ weaken linearly with $m_\chi$ in this limit. This relationship between the couplings and the reference cross section also explains why the 5th force constraints become more aggressive at higher $m_\chi$ for fixed $m_\phi$ (see Section~\ref{sec:other_constraints} and Ref.~\cite{Knapen:2017} for more discussion).

    For the fixed mediator case where $m_\chi < m_\phi / v_0 \sim 1~\rm{keV}$, the cross-section is expected to scale as:
    \begin{equation}
        \bar \sigma \propto 
        \frac{\sqrt{N_{\rm ind}}}{N^2}
        m_\chi^3r_C^2
        \quad \rm{for~}m_\phi = \text{constant} \; .
    \end{equation}

    \item There is a third limit for $m_\chi \propto m_\phi$ that lies in between the first two limits: when $q < r_C^{-1}$ and $q\sim (\Delta x)^{-1}$. Since $q < r_C^{-1}$, the rate still has an $N^2$ enhancement; however, since $q \sim (\Delta x)^{-1}$, the probability of producing decoherence is also high. In this intermediate regime, which is between the two ``knees", the two mediator cases behave differently.

    For $m_\phi \propto m_\chi$, $q < r_C^{-1}$ boosts the $m_\chi^{-3}$ dependence of the rate, which dominates with respect to the $m_\chi^{-1}$ discussed in the second limit. Thus, we expect:
      \begin{equation}
        \bar{\sigma} \propto 
        \frac{\sqrt{N_{\rm ind}}}{N^2} 
        r_C^2R_{\phi \chi}^4
        m_{\chi}^3, 
    \end{equation}
where $R_{\phi \chi} = m_\phi / m_\chi$ is the constant ratio assumed between the mediator and the dark matter masses.
\end{enumerate}

The {\it sweet spot} for sensitivity of the matter interferometers occurs in the threshold between limit 1 and 3, corresponding to a change in the sign of the slope in the curves in Fig.~\ref{fig:results}. These two limits overlap at the $m_\chi$ with best sensitivity, where the rate is boosted by both a large probability of decoherence and the $N^2$ Born enhancement. Equating both limits we expect the best constraint for a given experiment to happen at:
\begin{equation}
    m_\chi^\text{knee} \sim \frac{1}{R_{\phi \chi} \sqrt{\Delta x \, r_C}}.
\end{equation}

\subsubsection{Phase}
The limits of this subsection apply to both matter interferometers and cold atomic clouds.  As mentioned above, the phase effect relies on anistropy of the DM flux. Thus, we must include the Earth's velocity, $v_e$, when calculating the phase effect. This gives a phase rate of:
\begin{align}
\begin{split}
    R_{\rm{phase}} = \frac{1}{\rho_TV}\frac{\rho_\chi}{m_\chi} \frac{\pi \bar{\sigma}v_0^2 (q_0^2 + m_{\phi}^2 )^2}{2N_0m_{\chi}^2m_{\phi}^4}\int dq \frac{q}{\left((q/m_{\phi})^2 + 1\right)^2} N[1 + AF_A^2(q) + A(N_A - 1)F^2(qr_C)]\\
    \times \int_{-1}^1 d\cos\theta \left(-\sin(\mathbf{q}\cdot\mathbf{\Delta x})\right)\left[\exp\left(-\frac{v_-^2(q,\theta)}{v_0^2}\right) - \exp\left(- \frac{v_{\rm esc}^2}{v_0^2}\right)\right].
    \end{split}
    \label{eq:Rphase}
\end{align}
Since the presence of the $v_e$ term in Eqn.~\ref{eqn:vmin}~promotes the integrand to an odd function, the dark matter flux anisotropy is critical for obtaining a non-zero rate.  

We now analyze the behavior of the rate in the limit of different DM masses. As in the decoherence case, we discuss limits in the context of two mediator scenarios, shown in Figures~\ref{fig:results} \& \ref{fig:results_fixedmphi}: where $m_\phi$ is some constant ratio of $m_\chi$, and where $m_\phi$ is fixed to some value. 

\begin{enumerate}
    \item $m_{\chi} \rightarrow 0$ -- As before, in this limit $q \, r_C \ll 1$ and $q \, \Delta x\ll 1$. The first of these makes the rate go as $N^2$, and the second of these allows us to expand $\sin (\mathbf{q} \cdot \mathbf{\Delta x}) \simeq q \, \Delta x  \cos \theta$ (to first order in the Taylor expansion) in Eqn.~\ref{eq:Rphase}. Hence, the cross section constraints in this limit scale as:
        \begin{equation}
            \lim_{m_\chi \to 0} \bar{\sigma} \propto 
            \frac{\sqrt{N_{\rm ind}}}{N^2} 
        \Delta x \; .
        \end{equation}
        This behaviour applies independently of the $m_\phi$ scenario. This is because, as before, $\lim_{m_\chi \to 0}{\cal F}_\text{med} \sim \text{constant in }m_\chi$ in both cases.
        Note that the sensitivity of the phase measurement in this limit is expected to be \textit{constant}. This constant shift in the phase induced by the low DM momentum transfer can be understood as the classical effect of having a wave with large wavelength acting on two clouds separated by a physical distance of $\Delta x$. We only see this constant behavior at the lower $m_\phi$ ratios in Fig.~\ref{fig:results} because the $m_\phi/m_\chi$ ratio sets the intercept with the next limit case (see below).
    \item $m_\chi \to \infty$ -- $R_\text{phase} \to 0$ when $q \, \Delta x \gg 1$ because of the oscillatory behavior of the sinusoidal function (see right panel of Fig.~\ref{fig:cartoon} for illustration). Thus, we are interested in the limit where $m_\chi \to \infty$, but $v \ll v_0$. This implies that $q \, \Delta x \lesssim 1$ will bound the growth of $q$, which will then be decoupled from $m_\chi$. Furthermore, because we are now in the $q \, \Delta x \lesssim 1$ regime, we may expect the $N^2$ term in the target form factor will dominate (since for most of the missions $r_C \sim \Delta x$). Hence, we expect the cross section constraint to scale as:
        \begin{equation}
          \lim_{m_\chi \to \infty} \bar{\sigma} \propto
           \frac{\sqrt{N_{\rm ind}}}{N^2}
        \frac{(\Delta x)^3 \, m_{\chi}^4}{r_{\phi \chi}^4} \quad \text{ for }m_\phi \propto m_\chi\; ,
        \end{equation}
        where an extra $1/(m_\chi \Delta x)$ in the rate comes from the phase integral, as Eqn.~\ref{eqn:phase_effect} explicitly shows\footnote{Notice that $\tilde v_e$ factorizes out in the right-hand side of Eqn.~\ref{eqn:phase_effect} due to the expansion of the hyperbolic sine and it scales inversely proportional to $m_\chi \Delta x$.}. The $\bar{\sigma}_n \propto m_\chi^{4}$ scaling determines the slope of the curves in the logarithmic plots of Fig.~\ref{fig:results}. The intercept is fixed by $\Delta x$, $N$, and the ratio $r_{\phi \chi} = m_\phi / m_\chi$.

        On the other hand, when $m_\phi$ is fixed, the cross section sensitivity scales as:
        \begin{equation}
          \lim_{m_\chi \to \infty} \bar{\sigma} \propto
           \frac{\sqrt{N_{\rm ind}}}{N^2} 
        (\Delta x)^3 \, m_{\phi}^4
          \quad \text{ for a fixed }m_\phi\;,
        \end{equation}
       which is independent of $m_\chi$ (as long as this limit holds). If $m_\phi < 1/\Delta x$, the constant scaling with $m_\chi$ would go as $\bar \sigma \propto \sqrt{N_{\rm ind}}/(N^2 \Delta x)$.
\end{enumerate}

We note that the phase shifts discussed here can only be seen if the visibility is high enough to measure the fringe reliably. Thus, there cannot be a large decoherence effect. The phase limits we show in our figures are then only valid when they are more constraining than the decoherence limits. We include the full curves as a reference.

One final note before we move on to other interactions: if the DM scatters off of the \textit{electrons} in the clouds instead, we would get a similar rate: this rate may also get an $N^2$ enhancement since the electrons in these experiments are also coherent. However, we do not explicitly include these results here because the limits from astrophysical constraints are much more constraining for electron couplings \citep[see, e.g.,][]{Knapen:2017}.

\subsection{Hidden Photon Processes and Coherent Axion Scattering}\label{sec:other_process}

In this section, we consider other processes that could cause decoherence and phase shifts in an atom interferometer through coherently scattering with one of the two wavepackets. We discuss three different processes:~DM~scattering mediated by a hidden photon through kinematic mixing, hidden photon scattering through baryon and lepton number couplings, and coherent axion scattering. The hidden photon kinematic mixing scenario has a rate suppressed by the poor polarizability of the atom clouds in these experiments. The rates for the other two processes can be projected onto the rate for nuclear recoil with heavy mediators, and thus are not optimized for detection. Overall, these processes do not prove as compelling for atom interferometers as the nuclear recoil with light mediators. We focus on the decoherence observable within matter interferometers (Pino, and MAQRO) only in this section; however, the phase shifts can be calculated using the methods of the previous section.

\begin{itemize}

\item Hidden photon through kinetic mixing

A hidden photon, $A'$, couples to the SM photon through a kinetic mixing operator ${\cal L} \supset \frac{\kappa}{2} F^{\mu\nu}F'_{\mu\nu} $, where $\kappa$ is the kinetic mixing parameter, $F_{\mu\nu}$ is the photon field strength, and $F'_{\mu\nu}$ is that for the dark photon. The SM photons polarize the atom clouds with polarizability $\alpha$, and the effective Hamiltonian for the interaction between the dark photon, the SM photon, and the target medium then reads: %
\begin{equation} \label{eq:HintA}
H_{\rm I} = -  \alpha \, \kappa \int d^3 \mathbf{r}\, n(\mathbf{r}) \mathbf{E} (\mathbf{r}) \cdot \mathbf{E}' (\mathbf{r}) \; ,
\end{equation}
where $\mathbf{E}$ is the electric field for the SM photons, $\mathbf{E}'$ is the electric field for the dark photon and $n(\mathbf{r})$ is the number density of the solid.

With this interaction Hamiltonian, we can calculate the overall decoherence for DM scattering mediated by the hidden photon in the absence of external fields using the real part of Eqn.~\ref{eqn:overall_dec}:
% \CM{Cite Kathryn's paper}
\begin{equation}\label{eqn:dec_photon_pol}
    s =
    \frac{1}{N_{\rm ind}}
   % ( \alpha n_0 )^2
    \left( \frac{\alpha}{ a_0^3 }\right)^2
    \frac{\rho_\chi}{m_\chi} \frac{\pi \bar{\sigma}}{\mu^2} t_{\rm{exp}}\int \frac{d^3\mathbf{q}}{(2\pi)^3} \mathcal{F}_{\rm{med}}^2  \left(1 - \cos(\mathbf{q}\cdot\mathbf{\Delta x})\right) \sum_f |\langle f | \mathcal{F}_T(\mathbf{q}) | i \rangle |^2 g(\mathbf{q}, \omega_1)\; ,
\end{equation}
 where
 %$n_0$ is the average number density
 $a_0$ is the Bohr radius. Following the formalism in \cite{Knapen2017b}, we define the reference cross section $\bar{\sigma}$ as:
\begin{equation}\label{eqn:ref_cross_photon_pol}
\begin{gathered}
\includegraphics[width=0.17\textwidth]{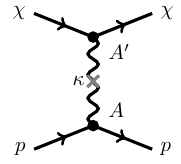}
\end{gathered} \quad : \quad
% \bar{\sigma} = \frac{\kappa^2}{2\pi} \frac{\mu^2 }{m_{A'}^2} \frac{m_{\chi}^2 v_{\rm DM}^2}{(m_{\chi}^2 v_{\rm DM}^2 + m_{A'}^2 )^2} \; .
\begin{gathered}
\bar{\sigma} = \kappa^2 \frac{\mu^2}{\pi}  \frac{1}{(m_{\chi}^2 v_{0}^2 + m_{A'}^2 )^2} \; ,
\end{gathered}
\end{equation}
where we have assumed the DM charge under the hidden photon to be 1. We use the mediator form factor:
\begin{equation}\label{eqn:mediator_photon_pol}
\mathcal{F}_{\rm{med}} (q) = \frac{(m_{\chi} v_{0})^2 + m_{A'}^2}{q^2 + m_{A'}^2 } ,
\end{equation}
and the structure function (convoluted with the kinematic function):
\begin{equation}\label{eqn:darkphotonS}
  \sum_f |\langle f | \mathcal{F}_T(\mathbf{q}) | i \rangle |^2 g(\mathbf{q}, \omega_1)
  = a_0^6 \frac{q^2}{2} \int \frac{d^3 \mathbf{k}_1}{(2\pi)^3} \omega_1 S(\mathbf{q} - \mathbf{k}_1)g(\mathbf{q}, \omega_1)
  \; ,
\end{equation}
where the static structure factor $S(\mathbf{q} - \mathbf{k}_1)  = \sum_{i,j=1,...,N_A} \langle e^{-i\Delta \mathbf{y}_{ij}\cdot (\mathbf{q} - \mathbf{k}_1)} \rangle_{\rm{target}}$ follows the same form as the one derived for nuclear recoil in Eqn.~\ref{eq:S} (see Sec.~\ref{sec:appendix1} for a detailed derivation). Notice that $\omega_1$ is the energy of an on-shell photon and obeys the dispersion relation $\omega_1 \approx k_1$.

We perform a numerical evaluation of the decoherence effect with $ \alpha /a_0^3 \sim \mathcal{O}(100)$, and find that the limits are $\gtrsim$ 20 orders of magnitude lower than the nuclear recoil limits in the light $m_\chi$ regime.

\item

Hidden photon through baryon and lepton number couplings

A hidden photon can couple coherently to the total baryon number of the atom cloud through:
\begin{equation}
    \mathcal{L}\supset \frac{g_B}{3}\Bar{q}\gamma_\mu q Z_B^\mu,
\end{equation}
where $g_B$ is the gauge coupling constant.

The rate for $Z_B$ scattering, and thus the overall decoherence, mediated by a nucleon, has the same dynamical structure
function as nuclear recoil given by~Eqn.~\ref{eq:S}.
 However, the reference cross section in this case is defined as:
\begin{equation}
\begin{gathered}
    \includegraphics[width=0.2\textwidth]{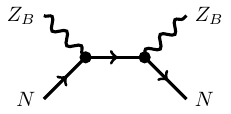}
    \end{gathered} : \quad \bar \sigma = \frac{3}{16\pi m_N^2}\left(\frac{g_B}{3}\right)^4 \; ,
    \label{eqn:DiagramZB}
\end{equation}
with a trivial mediator form factor $\mathcal{F}_{\rm{med}} (q) = 1$. We note that this cross section scaling is similar to that of nuclear recoil mediated by a heavy mediator. This is a consequence of the vector nature of the baryonic force on quarks; if a non-zero axial interaction is present, the cross section would be enhanced by a factor $m_{N}^2/m_{Z_B}^2$. The sensitivity curves for a heavy mediator are shown in Fig.~\ref{fig:heavy_reach}, which can be converted to limits on $g_B$ using the reference cross section above.

Similarly to the baryon number coupling scenario, if the hidden photon couples to the net lepton number, the rate would also benefit from the same coherent scattering enhancement. However, as mentioned at the end of Sec.~\ref{sec:nuc_recoils}, the electronic interactions of DM are more constrained that those involving baryons~\cite{Knapen:2017}.

\item Coherent axion scattering

Generally, couplings between axion DM and fermions are spin-dependent, which would break the coherence within one wavepacket and suppress the scattering rate. However, as has been recently pointed out in Ref.~\cite{Fukuda:2021drn}, there is a spin-independent interaction between axions and nucleons:
\begin{equation}
    \mathcal{L}\supset \frac{a^2}{8f_a^2} \sum_{N=p,n} \delta m_N \, \bar{N} N \; ,
\end{equation}
where $a$ is the axion field, $f_a$ is the axion decay constant, and the mass parameter $\delta m_N \sim \mathcal{O} (10)$ MeV characterizes the quark mass contribution to the nucleon mass. Importantly, such an interaction is not additionally suppressed by the axion mass $m_a$, as is normally expected from the expansion of the axion-nucleon interacting term. Instead, it is proportional to the quark masses, encoded in $\delta m_N$, which also violate the shift-symmetry in the UV Lagrangian.
This spin-independent interaction, combined with the large occupation number of axion DM, at first seems like a promising candidate for detecting axion DM with atom interferometers.

This effective interaction produces coherent axion scattering with the atom cloud, which causes an overall decoherence similar to the nuclear recoil case with heavy mediators. The reference cross section in this case is then:
\begin{equation}
   \bar{\sigma} = \frac{\mu^2}{256\pi}{\left(\frac{\delta m_N}{f_a^2}\right)}^2 \frac{1}{m_a^2} \; ,
\end{equation}
with a trivial mediator form factor. Recasting the heavy mediator limit shown in Fig.~\ref{fig:heavy_reach}, this probes $f_a$ only up to $100 \text{ GeV}$ and thus is not a promising target.

\end{itemize}

\section{Comparison with Other Constraints}\label{sec:other_constraints}

In this section, we put the limits we derive above into context. We first describe how they compare to previous atom interferometer dark matter limits. Then, we compare our results to other direct detection methods, and relevant astrophysical and cosmological constraints.

\subsection{Comparison to previous calculations}
Refs.~\citep{Riedel2013, Riedel2017} were the first to suggest using atom interferometers to probe dark matter through the decoherence effect we consider in this paper. They solely consider nuclear recoil calculations and their results are equivalent to ours if we choose the same statistical methods from each paper. Ref.~\citep{Riedel2017} only considers the decoherence observable. While \citep{Riedel2013} does consider the phase observable, it is not clear how they set their statistics for the phase. Several other works \citep{Geraci2016, Graham2016, Arvanitaki2018} have considered the phase effects in atom interferometers from ultralight dark matter (ULDM). However, these works rely on phase effects produced by varying the fine structure constant and electron mass, which are quite distinct from the effects we consider here.

\subsection{Comparison to other direct detection methods}

Most direct detection experiments cannot probe the small DM masses to which atom interferometers will be sensitive. The notable exceptions will be superfluid Helium experiments \citep{Knapen2017b} and phonon excitations in semiconductors \citep{Griffin2020}, as shown by the grey lines in Fig.~\ref{fig:results}. These experiments will outperform the atom interferometers we discuss in this paper for mediator masses $m_\phi > 10^{-7}m_\chi$; however, none of these experiments can probe below $m_{\chi} \lesssim 10~\rm{keV}$. Another recent work has proposed using optomechanical sensors to measure DM scattering \citep{Afek2021}. The most ambitious of these would produce comparable limits in the heavy mediator case to MAQRO in the $0.01-1~\rm{MeV}$ range.

Another promising avenue for probing light DM is to use atomic clocks or magnetometers \citep{Alonso2019, Wolf2019}. Although the constraints on the scalar DM-nucleon cross sections from these experiments are much stronger, these experiments also have many more nuclei. The constraining power for the same number of nuclei and same integration time is similar between these experiments and atom interferometers. However, these atomic clock and magnetometer experiments must rely on spin-dependent DM interactions to produce observable effects. This does not allow them to gain the coherent enhancement that atom interferometers may use (depending on the type and the observable). If these experiments could probe spin-independent DM interactions and gain the $N^2$ enhancement, then the earlier spin-dependent results suggest that they could be more powerful than the atom interferometers, given the same nucleon numbers.

Atom interferometers are so far the only way to effectively probe $m_\chi \lesssim 10~\rm{keV}$ DM with spin-independent interactions. Note however that atom interferometers can \textit{only} effectively probe spin-independent DM interactions -- any spin-dependent effects will break the coherence that is used to get such effective limits in the observables we consider. Thus, some combination of atom interferometers and the other direct detection experiments described above are necessary to rule out this parameter space.

Although not expressly shown in the figures, we want to stress that the space missions we discuss in this paper would also be sensitive to large cross section DM. If DM had very large nucleon cross sections, then current, terrestrial direct detection experiments would not be able to detect DM due to shielding by the atmosphere. The space-based missions we describe here would be complementary to proposed balloon and low-Earth-based orbit satellites \citep{Emken2019}.

\subsection{Astrophysical \& Cosmological Constraints}
In Fig.~\ref{fig:results}, we show in gray the parameter space ruled out by the existing astrophysical bounds on the reference cross section described in Eqn.~\ref{eqn:RefXSec}. We briefly address these and other astrophysical and cosmological limits in this section, and refer the reader to \citep{Knapen:2017} for further details.

There are two couplings entering in the reference cross section, $\bar \sigma$:
\begin{enumerate}
    \item $y_n$, which quantifies the interaction between the mediator, $\phi$, and the nucleons. Light bosonic mediators can be emitted in stars, leading to their potential rapid cooling. Constraints on new forms of energy loss in stars then translate into a bound for the coupling $y_n$~\cite{Hardy:2016kme,Ishizuka:1989ts}. Particularly stringent for the light mediator masses that we are considering are the fifth force constraints. The most relevant bounds come from Casimir force, cantilever, and torsion balance experiments~\cite{Murata2015}, which constrain the attractive Yukawa potential generated by the $\phi$-nucleon interaction,
    \begin{equation}
        V(r) = -\frac{y_n^2}{4\pi} \frac{1}{r} e^{-m_\phi r},
    \end{equation}
    which translates into a bound on $y_n$ for a given $m_\phi$.
    The solid gray shaded area in Fig.~\ref{fig:results} corresponds to the parameter space in the $\bar \sigma$ - $m_\chi$ plane excluded by the above constraints when assuming a large $\phi$-$\chi$ coupling, $y_\chi = \sqrt{4\pi}$.
    \item $y_\chi$, which weights the interaction between the mediator, $\phi$, and the dark matter, $\chi$. If $\chi$ composes all of the DM in the universe, dark matter self-interactions (DMSI) are constrained by cluster mergers and halo shaped observations to satisfy $\sigma_\text{DMSI} / m_\chi < 1 - 10 \text{ cm}^2/\text{g}$~\cite{Bondarenko:2020mpf}, which together with the previous bounds on $y_n$, impose severe constraints on $\bar \sigma$. This would rule out the possibility of testing DM with atom interferometers in the foreseeable future. However, the DMSI bound can be considerably relaxed if $\chi$ is a subcomponent of the total DM. In this paper, we assume that $\chi$ composes $5\%$ of the total DM.
    \end{enumerate}
For the light DM particles we are considering, cosmological observations can also impose relevant bounds on the couplings $y_\chi$ and $y_\phi$, and thus indirectly on $\bar \sigma$, by constraining the contribution of the light particles from the dark sector to the number of effective relativistic degrees of freedom ($N_\text{eff}$). The contribution of distinct relativistic fields from the dark sector (each denoted with subscript $i$) to $N_\text{eff}$ is given by:
\begin{equation}
    \Delta N_\text{eff} = \frac{4}{7} \sum_i  \, g_i \, \left(\frac{g(T_{\nu_L}^\text{dec})}{g(T_i)}\right)^{4/3},
\end{equation}
where $g_i$ are the number of degrees of freedom of the particle $i$, $T_{\nu_L}^\text{dec}$ is the decoupling temperature of the active neutrinos, and $T_i$ is the temperature of the dark sector.

Whatever UV interaction that generates the coupling of the mediator with the nucleons ($y_n$) will also induce an effective interaction between $\phi$ and the gluons. Then, above the QCD phase-transition, thermal scatterings of the mediator with gluons, such as $gg \to \phi g$, can bring the mediator into thermal equilibrium with the SM bath. Once thermalized, the mediator will decouple as the universe cools down, depending on the strength of $y_n$.

For a light mediator ($m_\phi \propto m_\chi$ or fixed $m_\phi = 1 \text{ eV}$), for all the parameter space that we consider, $y_n < 10^{-9}$ because of the stringent $5^\text{th}$ force and stellar constraints. In this case, $\phi$ decouples from the SM bath before the QCD phase transition, as derived in Ref.~\citep{Knapen:2017}. After the decoupling of $\phi$, the dark matter $\chi$ cannot be thermally produced and also decouples from the SM bath. The contribution to the number of relativistic degrees of freedom from the dark sector then depends on the number of relativistic particles that were in thermal equilibrium at the time the dark sector decouples. Hence, $\Delta N_\text{eff} \simeq  0.06 \sum_i  g_i$, where $g_i$ depends on the nature of the dark matter and the mediator. Note that such a small contribution is far from being in tension with the current $N_\text{eff}$ measurements, but will be easily probed by upcoming experiments, such as CMB-S4~\cite{Abazajian:2019eic}, PICO~\cite{NASAPICO:2019thw}, CORE~\cite{CORE:2016npo}, and CMB-HD~\cite{CMB-HD:2022bsz}.

Finally, we want to address the particular range in $m_\chi$ that we choose for our plots. The upper limit is set by $m_\chi \ll m_N$, which allows us to make several simplifying assumptions in the nuclear recoil case. Normally, the lower bound would be set by warm DM constraints (typically at $m_\chi \sim 1~\rm{keV}$, see for example Ref.~\cite{Gilman2020}); however, we consider $\chi$ to be a small subcomponent of the DM. Thus, we set our lower bound by mandating that the DM velocity should be less than the escape velocity of the Milky Way. This sets the reasonable limit that we should have DM to detect in our galaxy. If, as we discuss above, the DM decouples above the QCD phase transition, this sets a temperature for the DM of $T_{\rm{DM}} \sim 0.1 T_{\gamma} \sim 10^{-5}~\rm{eV}$ today, where $T_{\gamma}$ is the temperature of the cosmic microwave background today. Approximating $v \sim \sqrt{T_{\rm{DM}}/m_\chi}$ and setting $v \lesssim 0.001$ gives a lower limit on the DM mass of $m_\chi \gtrsim 10 \mbox{ eV}$.

% Finally, some comments about the depletion of the $\chi$ and $\phi$ relic abundances are in order. Following~\citep{Knapen:2017}, we could consider an extra light degree of freedom, $a$, with mass $m_a \ll \text{ eV}$, to which $\chi$ and $\phi$ can annihilate (or decay in case of $\phi$) consistently with the cosmological bounds~\citep{Knapen:2017}.

\section{Discussion \& Conclusions}\label{sec:conclusion}
In this paper, we consider the effects of dark matter scattering on atom interferometers. We calculate the decoherence and phase shifts produced in atom interferometers from several  different channels: nuclear recoils, hidden photon processes, and coherent axion scattering. The resulting forecasted limits for various proposed atom interferometer experiments are given in Figs.~\ref{fig:results}, \ref{fig:results_fixedmphi}, and \ref{fig:heavy_reach}, for the observables that are Born enhanced at momentum transfers below the inverse size of the cloud. In this final section, we discuss some caveats and opportunities for improvement, both theoretically and experimentally.

We make various assumptions and approximations in this paper. For example, we neglect to include the mass of the targets in the kinematic matching equations (\textit{e.g.,} Eqn.~\ref{eqn:vmin}). However this will only have an effect at $m_\chi \sim m_N$, which is at the very upper end of the masses we consider here. In addition, we do not take into account certain detection mechanisms that could also affect the observable visibility change from DM. As one example: if the DM momentum transfer is large enough to move one of the clouds outside of the detection region, this would lead to extra visibility loss. However, we note that DM masses greater than our range of interest ($m_\chi >10~\rm{GeV}$) would be required to move a cloud this far. Finally, we provide a simplistic statistical formalism and do not include the DM flux modulation for setting the decoherence limits (although a discussion of this effect is included in Appendix~\ref{sec:appendix2}). For all of these cases, we have taken the more conservative approach to our approximations. One final caveat we should note: we do not consider any specific backgrounds that could affect these results. A recent study showed that long-distance forces are negligible for decoherence, especially for the short ($\lesssim 10$~s) measurement timescales we consider here \citep{Kunjummen2022}. As we discuss in Section~\ref{sec:overview}, we expect these processes to be suppressed relative to the light DM models with light mediators. However, for the space-based experiments we consider, we plan to carry out a careful study of the cosmic ray and solar photon backgrounds in the future.

Future work could address the above simplifications in more detail. Exploring how atom interferometers could probe other models, such as large bound states of DM (\textit{i.e.,} DM nuggets) could also be interesting \citep{Coskuner2019}. These models are also coherent at low momentum transfer, making them prime targets for atom inteferometers. Another avenue for probing DM could examine the collective excitations produced in Bose-Einstein condensates. While these could not be probed by traditional atom interferometers, other BEC experiments could be used.

In this paper, we only considered a few of the many proposed atom interferometer experiments \citep[see, e.g.,][]{El-Neaj2020, Badurina2020, Aguilera2014}. The experiments we chose have a wide spread in $\Delta x$, $N$, $r_C$, and $t_{\rm{exp}}$, so that our results span much of the parameter space that other interferometers could probe. %Our public code is flexible and can accommodate any single atom interferometer, as long as the parameters in Table~\ref{tab:mission_params} are known for a given experiment.
We note that in this paper we only consider how a single atom interferometer could probe DM; however, some of the experiment concepts, such as GDM, MAQRO and AEDGE, will have networks of atom interferometers, and will be especially sensitive to differential phase measurements between two atom interferometers, potentially improving the reach. We leave a detailed investigation of atom interferometer networks to future work.

We would now like to address how atom interferometers could improve their bounds on DM physics. As we discuss in Section~\ref{sec:nuc_recoils}, we find that the best constraint for matter interferometers can be approximated by: $m_\chi \sim (\sqrt{\Delta x \, r_C} \,  m_\phi / m_\chi)^{-1}$, if $\Delta x$ is of the same order or larger than $r_C$. Thus, experiments can target specific mass ranges by properly tuning the cloud separation and size. To probe the weakest part of the current astrophysical constraints, $10~\rm{eV}\lesssim m_{\chi} \lesssim 1~\rm{MeV}$ with $m_\phi = 10^{-5}~m_\chi$, these experiments would want $1~\rm{cm} \gtrsim \Delta x  \gtrsim 100~\rm{nm}$, with similarly sized cloud radii. The cross section limit sensitivity for atom interferometers is then mostly set by the number of nucleons -- the more, the better. The most important design change that these experiments could consider is to increase their nucleon count.

As we show in this paper, atom interferometers would be a complementary probe to current direct detection efforts. They would uniquely probe light dark matter, including improving current astrophysical bounds by up to 10 orders of magnitude in some cases. These and other quantum sensing missions should continue to be studied for use as DM probes.

\section*{Acknowledgements} The authors would like to thank the anonymous referee for their helpful comments.
The authors would also like to thank Sheng-wey Chiow, Curt Cutler, Ryan Plestid, Marianna Safronova, and Tanner Trickle for useful discussions. Part of this work was done at the Jet Propulsion Laboratory, California Institute of Technology, under a contract with the National Aeronautics and Space Administration. The work of KZ is supported by the DoE under contract DE-SC0011632, and by a Simons Investigator award. This work is also supported by the Walker Burke Institute for Theoretical Physics.

 \textit{Software:} astropy \cite{astropy}, matplotlib \cite{matplotlib}, numpy \cite{numpy}, PhonoDark \cite{Trickle2020a}, scipy \cite{scipy}

\appendix

\section{Daily Modulation}\label{sec:appendix2}

As has been shown in the main text, e.g.~through~Eqn.~\ref{eq:pdec}, both the decoherence rate and the phase are sensitive to the orientation of $\mathbf{\Delta x}$. This directional signal will modulate during a day (orbit period), as the Earth (space-based experiment), rotates around the Earth's axis (Sun). In this appendix, we present a daily modulation formalism and calculation to show the significance of the time-varying signal. For space-based experiments, the modulation will depend explicitly on the orbit. We expect this to have a similarly-sized effect as the daily modulation, since they originate from the same directional information of $\mathbf{\Delta x}$. We leave a specific orbit study to future work.

% Note that we ignore the yearly modulation as the~Earth orbits around the sun, due to its relatively small effect compared to the daily modulation.

Daily modulation effects can be parameterized through the time-dependence of the Earth's velocity $\mathbf{v}_e(t)$ in the lab frame, where the phase-space distribution of the dark matter takes the boosted Maxwell-Boltzmann distribution as in~Eqn.~\ref{eq:fv}. In the lab frame where ${\bf \Delta x}$ is chosen to be aligned with the ${\bf \hat{z}}$ axis, the Earth's velocity at time $t$ reads:
\begin{equation}
     {\bf v_\text{e}}(t) = \|\mathbf{v}_e\| \begin{pmatrix}
    \text{s} \theta_\text{e}  \, \text{c} \theta_\text{x} \, \text{s} \phi(t)- \text{s} \theta_\text{x} \, \text{s}  \theta_\text{e} \, \text{c} \theta_\text{l} \, \text{c} \phi(t)+ \text{s} \theta_\text{x} \, \text{c} \theta_\text{e}  \, \text{s} \theta_\text{l} \\

 \text{c} \theta_\text{e} \, \text{s} \theta_\text{g} \,  \text{s} \theta_\text{l} \, \text{c} \theta_\text{x}  - \text{s} \theta_\text{e} \, \text{s} \theta_\text{g} \, \text{c} \theta_\text{l} \, \text{c} \theta_\text{x} \, \text{c} \phi(t) -
 \text{s} \theta_\text{e} \, \text{c} \theta_\text{g} \, \text{s} \theta_\text{l} \, \text{c} \phi(t) -
 \text{c} \theta_\text{e} \, \text{c} \theta_\text{g} \, \text{c} \theta_\text{l} - \text{s} \theta_\text{e} \, \text{s}\theta_\text{g} \, \text{s}\theta_\text{x} \, \text{s}\phi(t) \\
 \text{s}\theta_\text{e} \, \text{c}\theta_\text{g} \, \text{c}\theta_\text{l} \, \text{c}\theta_\text{x} \, \text{c}\phi(t)-\text{c}\theta_\text{e} \, \text{c}\theta_\text{g} \, \text{s}\theta_\text{l} \, \text{c}\theta_\text{x}-\text{s}\theta_\text{e} \, \text{s}\theta_\text{g} \, \text{s}\theta_\text{l} \, \text{c}\phi(t)-\text{c}\theta_\text{e} \, \text{s}\theta_\text{g} \, \text{c}\theta_\text{l} +\text{s}\theta_\text{e} \, \text{c}\theta_\text{g} \, \text{s}\theta_\text{x} \, \text{s} \phi(t)
     \end{pmatrix},
\end{equation}
where the abbreviations ``s" and ``c" refer to sine and cosine, respectively. In the above matrix, $\theta_\text{e}\approx 42^{\circ}$ is the angle between the Earth's velocity ${\bf v_\text{e}}(t=0)$ and the north pole, $\theta_\text{l}$ is the angle between the location of the experiment and the north pole, $\phi(t) =  2 \pi \times t/$24h accounts for the rotation of the Earth and introduces the time dependence, $\theta_\text{x}$ represents the orientation of ${\bf \Delta x}$ in the plane perpendicular to the free-falling direction $\mathbf{ \hat{g}}$, while the angle $\theta_\text{g}$ is the angle between ${\bf \Delta x}$ and such plane.

\begin{figure}[h]
     \centering
   \includegraphics[width=\textwidth]{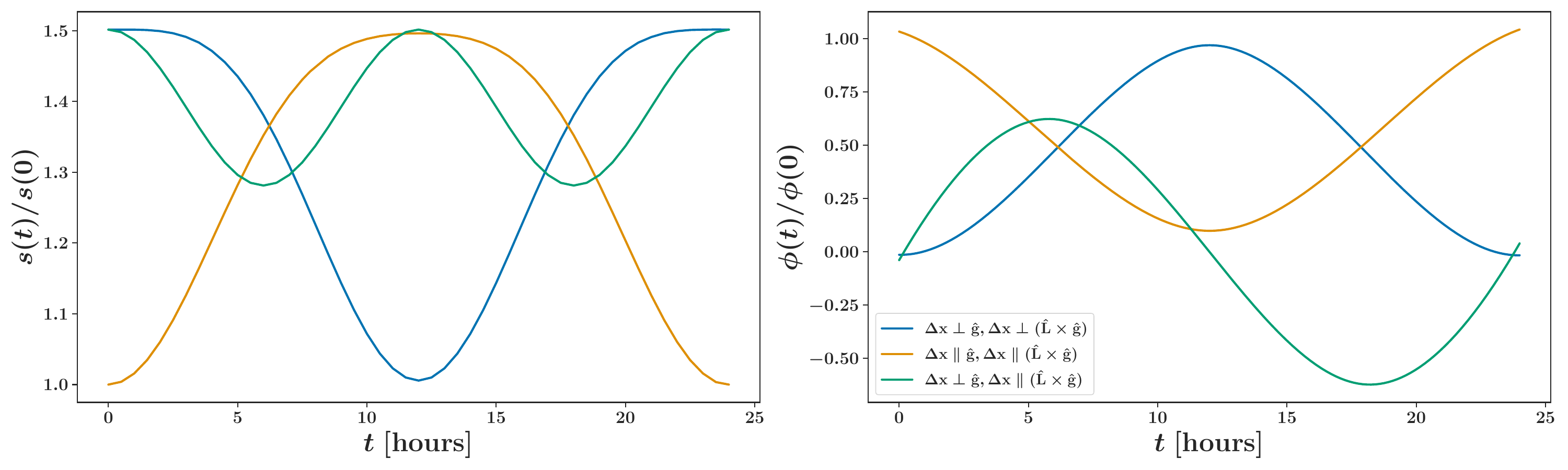}
   \caption{Daily modulation of the decoherence and phase signals for Pino, at latitude $47^{\circ}$N, assuming three different orientations of $\mathbf{\Delta x}$: $\mathbf{\Delta x} \perp {\bf \hat{g}}$ and $\mathbf{\Delta x} \perp ({\bf \hat{L}} \times {\bf \hat{g}})$ (blue); $\mathbf{\Delta x} \perp {\bf \hat{g}}$ and $\mathbf{\Delta x} \parallel ({\bf \hat{L}} \times {\bf \hat{g}})$ (orange); $\mathbf{\Delta x} \parallel {\bf \hat{g}}$ (green). ${\bf \hat{g}}$ is the direction of free-fall and ${\bf \hat{L}}$ is aligned with the Earth's spin axis ($\vec{\rm SN}$). Left panel shows the normalized decoherence effect $s(t)/s_0$ and the right panel shows the normalized phase effect $\phi(t)/\phi_0$. The normalization factor $s_0 = s_{\mathbf{\Delta x}\parallel {\bf \hat{g}}}(t=0)$ and $\phi_0 = \phi_{\mathbf{\Delta x}\parallel {\bf \hat{g}}}(t=0)$. We use $m_\chi = 1$~MeV and $m_\phi = 10^{-3} m_\chi$ for the plots.}
     \label{fig:dailyMod}
 \end{figure}

We show the daily modulation for Pino in~Fig.~\ref{fig:dailyMod}, taking three different orientations for $\mathbf{\Delta x}$. Although the final modulation is sensitive to the orientation, as well as the orbit information for space-based experiments, the different curves shown in Fig.~\ref{fig:dailyMod} give an estimate of the range of modulations we expect.
 By taking $\theta_\text{x} = 0$ and $\theta_\text{g} = 0$, we set ${\bf \Delta x} \perp {\bf \hat{g}}$ and ${\bf \Delta x} \perp ({\bf \hat{L}} \times {\bf \hat{g}})$, shown in the blue lines of Fig.~\ref{fig:dailyMod}, where ${\bf \hat{L}}$ is the unit vector for the Earth's spin axis ($\vv{\rm SN}$). Taking $\theta_\text{x} = \pi/2$ and $\theta_\text{g} = 0$, we are considering the experiment to be ${\bf \Delta x} \perp {\bf \hat{g}}$ and ${\bf \Delta x} \parallel ({\bf \hat{L}} \times {\bf \hat{g}})$, represented with orange lines in the same figure. Finally, taking $\theta_\text{g} = \pi/2$ we are considering that the experiment is (anti)-aligned with the free-falling direction, represented by the green lines in Fig.~\ref{fig:dailyMod}.

 Notice that for large~DM mass, where $m_{\chi} v_0 \gg 1/\|\mathbf{\Delta x}\|$, there is no daily modulation. In the limit of low~DM mass, where $m_{\chi} v_0 \ll 1/\|\mathbf{\Delta x}\|$, the amount of daily modulation does not depend on the DM mass. However, the total decoherence factor is suppressed in this limit, as explained in the main text. Unexpectedly, the mediator mass $m_{\phi}$ does not play a major role in the daily modulation. Although large momentum transfer is preferred for heavy mediator mass $m_{\phi} \gg m_{\chi}$, the anisotropy of the target still picks up daily modulation in the low to medium DM mass range, similar to other directional detection scenarios \citep{Coskuner:2021qxo}.
 Overall, we estimate the daily/orbit modulation to affect the signal by a factor of a few.

% We define the lab frame as follows: the $y$-axis is defined to align with the free-fall direction $\hat{g}$, i.e.~$\hat{y} = \hat{g}$. Let the unit vector defining Earth's rotation axis be $\hat{L}$ (pointed in the $\vv{\rm SN}$ direction). The $x$-axis is defined to be perpendicular to the plane spanned by $\hat{g}$ and $\hat{L}$, i.e. $\hat{x} = \frac{\hat{L} \times \hat{g}}{\| \hat{L} \times \hat{g}\|}$. The $z$-axis is defined accordingly for a right-handed coordinate system. In the lab frame defined above, the Earth's velocity reads:
% \begin{equation} \label{eq:vet}
% \begin{split}
% \mathbf{v}_e(t) = \|\mathbf{v}_e\|
% \begin{pmatrix}
% -\sin \phi_e \sin \theta_e \\
% -\cos \theta_e \cos\theta_l -\cos\phi_e\sin\theta_e\sin\theta_l\\
% \cos\phi_e\cos\theta_l\sin\theta_e-\cos\theta_e\sin\theta_l
% \end{pmatrix},
% \end{split}
% \end{equation}
% with $\phi_e = 2 \pi \times t/$24h parameterizing the~Earth's spin, $\theta_e \approx 42^{\circ}  $ defining the angle between the Earth’s rotation axis and the direction of its velocity, and $\theta_l$ giving the angle between the Earth’s rotation axis and the lab's direction. Notice that at $t=0$, $\mathbf{v}_e$ is on the plane spanned by $\hat{g}$ and $\hat{L}$. We let $\mathbf{\Delta x}$ be free to have any orientation in the lab frame, which results in different daily modulation signals.

\section{Derivation of Bose Einstein Condensate Form Factor \& Born Enhancement}
%%%%%%%%%%%%%%%%%%%%%%%
\label{sec:bec_appendix}

In this appendix we compute the form factor relevant for coherence amongst the $N$ nuclei in a Bose Einstein Condensate (BEC) built from optical or magnetic trapping. We mainly follow the discussion about coherency and incoherency from Ref.~\cite{Bednyakov_2018}.

The BEC, composed of $N$ nucleons, is initially in the ground state $(0)$. The probability of scattering with the BEC is given by the squared amplitude:
\begin{equation}
    |{\cal A}|^2  = \sum_m |{\cal A}_{m0}|^2 = |{\cal A}_0|^2   \sum_{k,j} \sum_m f_{m0}^k f_{m0}^{j*} \; ,
\label{eq:Amplitude}
\end{equation}
where we sum over all possible final states of the BEC after the scattering. The form factors $f_\text{m0}^\ell$ are defined as
\begin{equation}
\begin{split}
    f_{m0}^\ell(\mathbf{q}) &= \langle m |e^{i\mathbf{q \cdot \hat X_\ell}}|0\rangle = \int \left( \displaystyle \prod_{i=1}^N d \mathbf{x}_i\right)\Psi_\text{m}^* (\mathbf{x}_1, ..., \mathbf{x}_N) \Psi_0 (\mathbf{x}_1,..., \mathbf{x}_N)e^{i \mathbf{q}\cdot \mathbf{x_\ell}} \; ,
\end{split}
\end{equation}
where $x_i$ is the position of the $i$th nucleon and $\mathbf{\hat{X}}_i$ is the quantum position operator for the $i$th nucleon.

Summing over all possible final states (Parseval's identity),  Eq.~\eqref{eq:Amplitude} can be rewritten in the following way,
\begin{equation}
   | {\cal A}|^2 = |{\cal A}_0|^2 \sum_{k,j} \langle 0 | e^{-i \mathbf{q} \cdot \mathbf{\hat X}_j}|0 \rangle \langle 0 | e^{-i \mathbf{q} \cdot \mathbf{\hat X}_k}|0 \rangle + |{\cal A}_0|^2\sum_{k,j}\sum_{m\neq 0}\langle 0 | e^{-i\mathbf{q \cdot \hat X}_j}|m\rangle \langle m| e^{i \mathbf{q \cdot \hat X}_k}|0\rangle.
\end{equation}
Given that $f_{00}^k(\mathbf{q})$ does not depend on the index $k$ because of the symmetry properties of the wavefunction under the interchange of bosons, the first term of the above equation gives:
\begin{equation}
   |{\cal A}_0|^2 N^2 |F(\mathbf{q})|^2 \; ,
\end{equation}
which encodes the $N^2$ coherent enhancement. The form factor $F(\mathbf{q})$ quantifies coherence as a function of the properties of the BEC. It is defined as:
\begin{equation}
F(\mathbf{q}) = \langle 0| e^{-i \mathbf{q \cdot \hat X}_\ell}|0\rangle = \int \left(\prod_{i\neq \ell}^N d\mathbf{x}_i  |\Psi_0(\mathbf{x}_i)|^2 \right) d \mathbf{x}_\ell |\Psi_0(\mathbf{x}_\ell)|^2 e^{i\mathbf{q  \cdot x}_\ell} = \int d \mathbf{x_\ell} |\Psi_0 (\mathbf{x}_\ell)|^2 e^{i \mathbf{q \cdot x}_\ell} \; ,
\end{equation}
where we have used that the wave functions are normalized to unity and that the BEC is composed of non-interacting bosons $|\Psi_0(\mathbf{x}_1, ..., \mathbf{x}_N)|^2 = |\Psi_0(\mathbf{x}_1)|^2 ...|\Psi_0(\mathbf{x}_N)|^2$. Notice that the form factor $F(\mathbf{q})$ is just the Fourier transform of the density distribution (up to an $N$ normalization that we factorized out).

Let us assume that the BEC is built in a shallow trapping potential, which can be approximated to first order by the potential of a harmonic oscillator,
\begin{equation}
V_\text{BEC} \simeq \frac{1}{2} m \omega^2 r^2,
\end{equation}
where $m$ is the mass of the cloud, $\omega$ is the frequency of harmonic oscillator, and $r = (x^2 + y^2 + z^2)^{1/2}$. In the above equation we have also assumed that the trapping potential is spherical symmetric. The ground state wavefunction of such a potential is given by the following Gaussian:
\begin{equation}
\Psi_0(r) = \left(\frac{m\omega}{\pi \hbar}\right)^{3/4}  \text{exp}\left(-\frac{m}{2\hbar} \omega r^2\right) \; .
\end{equation}
The Fourier transform of $|\Psi_0(r)|^2$ is then given by:
\begin{equation}
    F(\mathbf{q}) = \text{exp}\left[-\left(\frac{q \, r_\text{cloud}}{2}\right)^2\right] \; ,
\end{equation}
where $r_\text{cloud}$, the radius of the cloud, is fixed by the width of the averaged width of the Gaussian (harmonic oscillator length) \; ,
\begin{equation}
    r_\text{cloud} = \left(\frac{\hbar}{m \omega}\right)^{1/2} \; .
\end{equation}
We will use the above form factor to parameterize the coherence $N^2$ enhancement in the phase-shifts of the missions that involve BECs.

% \CM{I don't see clearly how to cite the references (were not used to derive any of the above):}
% ...\citep{Javanainen1994, You1996}

\section{Derivation of the Hidden Photon Process through Kinematic Mixing}\label{sec:appendix1}

% \yw{Under construction. Do we still need to appendix?}

Here we derive the dynamic structure factor for the DM scattering mediated by a hidden photon through kinematic mixing, which we consider in Section~\ref{sec:other_process}. This largely follows the derivation in Ref.~\cite{Knapen2017b}. Consider a dark photon, $A'$, that couples to the SM photon through a kinetic mixing operator,
\begin{equation}
   {\cal L} \supset \frac{\kappa}{2} F^{\mu\nu}F'_{\mu\nu} \; ,
\end{equation}
where $\kappa$ is the kinetic mixing parameter, $F_{\mu\nu}$ is the photon field strength, and $F'_{\mu\nu}$ is that for the dark photon. The SM photon couples to the target material via its polarizability. To leading order, this gives a polarization $\mathbf{P} (\mathbf{r}) = \alpha \, n(\mathbf{r}) \mathbf{E} (\mathbf{r})$, where $\mathbf{E}$ is the total electric field in the medium, $n(\mathbf{r})$ is the number density of the medium. The polarizability can be normalized as $\alpha = (\varepsilon^{(r)} - 1)/n_0$, where $\varepsilon^{(r)}$ is the relative linear dielectric constant, and $n_0$ is the average number density. The Hamiltonian for the polarization then reads:
\begin{equation}
H_{\rm I} = - \frac{1}{2} \alpha \int d^3 \mathbf{r} \, n(\mathbf{r}) \mathbf{E} (\mathbf{r}) \cdot \mathbf{E} (\mathbf{r}) \; .
\end{equation}
After a field re-definition in the presence of a hidden photon, $A_{\mu} \rightarrow A_{\mu} + \kappa A'_{\mu}$, this gives
\begin{equation}
H_{\rm I} = -  \alpha \kappa \int d^3 \mathbf{r} \,  n(\mathbf{r}) \mathbf{E} (\mathbf{r}) \cdot \mathbf{E}' (\mathbf{r}),
\end{equation}
where $\mathbf{E}'$ is the electric field for the dark photon, that describes the interaction between the dark photon, the SM photon and the target medium.

In the atom interferometers that we consider in this paper, there are no external B-fields. Thus, we consider a scattering process mediated by a dark photon with mass $m_{A'}$ converted to a SM photon with momentum $\mathbf{k}_1$ and energy $\omega_1$. Following Ref.~\cite{Knapen2017b}, where the electric
field of the photon is quantized while the non-relativistic hidden photon field is sourced by a Coulomb
potential, the polarization-averaged squared matrix element of:
\begin{align}
    |\langle \mathbf{p}_i | H_I | \mathbf{p}_f;\mathbf{k}_1  \rangle |^2
    % |\langle \mathbf{p}_i | H_I | \mathbf{p}_f;\mathbf{k}_1; \mathbf{k}_2 \rangle |^2
    &= \left. \frac{\alpha^2\kappa^2}{2V^2} \omega_1
    \frac{q^2}{\left( q^2 + m_{A'} \right)^2}
    \left | \langle \Phi _0 | n_{-\mathbf{k}_2} | \Phi_0 \rangle \right|^2
     \right|_{\mathbf{k}_2 = \mathbf{q}- \mathbf{k}_1}
    % \delta_{\mathbf{q}, \mathbf{k}_1 + \mathbf{k}_2}
    \\
%     &= \left. \frac{\alpha^2\kappa^2}{2V} \omega_1
%     \frac{q^2}{\left( q^2 + m_{A'} \right)^2}
%   \frac{1}{V^2} \sum_{i,j = 1,\cdots, N_A }  \langle e^{-i \Delta \mathbf{y}_{i,j} \cdot \mathbf{k}_2}\rangle
% %   \frac{\delta_{\mathbf{q}, \mathbf{k}_1 + \mathbf{k}_2}}{V}
%      \right|_{\mathbf{k}_2 = \mathbf{q}- \mathbf{k}_1}
%   \\
    &=\left. \frac{\alpha^2\kappa^2}{2V^2} \omega_1
    \frac{q^2}{\left( q^2 + m_{A'} \right)^2}
   \frac{1}{V} S( \mathbf{k}_2 )
%   \frac{\delta_{\mathbf{q}, \mathbf{k}_1 + \mathbf{k}_2}}{V}
     \right|_{\mathbf{k}_2 = \mathbf{q}- \mathbf{k}_1}
   \\
    &= \left( \frac{\alpha}{ a_0^3 }\right)^2 \frac{1}{V^2} \frac{ \pi \bar{\sigma}}{\mu^2}  \mathcal{F}_{\rm med}^2(q) \;a_0^6 \frac{1}{V }  \omega_1
    \frac{q^2}{2 }
    S( \mathbf{q}- \mathbf{k}_1)
    % \delta_{\mathbf{q}, \mathbf{k}_1 + \mathbf{k}_2},
    % \delta(E_i - \omega_1 - E_f) \;
\end{align}
where $\mathbf{q}$ is the momentum transfer, $|\Phi_0\rangle$ is the ground state, and $n_{-\mathbf{k}_2}$ is the Fourier transform of the number density operator $n(\mathbf{r})$:
\begin{align}
  n_{-\mathbf{k}_2} = \frac{1}{\sqrt{V}} \sum_{i = 1,\cdots, N_A } e^{- i \mathbf{k}_2 \cdot \mathbf{y}_i},
\end{align}
and
the resulting static structure factor $S(\mathbf{k}_2) \equiv \sum_{i,j = 1,\cdots, N_A }  \langle e^{-i \Delta \mathbf{y}_{i,j} \cdot \mathbf{k}_2}\rangle$ has the same functional dependence as the one derived for the nuclear recoil case in Eqn.~\ref{eq:S} for contrast loss within matter interferometers and phase-shifts for both matter and diffuse atom cloud interferometers, which reflects that this process receives the same coherent scattering enhancement as the nuclear recoil case. Notice that in the last line of the derivation we have rearranged according to the reference cross section and mediator form factor, defined in Eqns.~\ref{eqn:ref_cross_photon_pol} and~\ref{eqn:mediator_photon_pol}. Factoring out the model-dependent dimensionless coupling $\alpha  n_0$, the model-independent structure function (convoluted with the kinematic function) for this process is:
\begin{equation}
    \begin{split}
\label{eqn:}
  \sum_f |\langle f | \mathcal{F}_T(\mathbf{q} ) | i \rangle |^2 g(\mathbf{q}, \omega_1)
  & = a_0^6 \frac{1}{V} \sum_{\mathbf{p}_f,\mathbf{k}_1} \omega_1  \frac{q^2}{2}  S( \mathbf{q}- \mathbf{k}_1 )g(\mathbf{q}, \omega_1) \\
  & = a_0^6 \frac{q^2}{2}  \int \frac{d^3 \mathbf{k}_1}{(2\pi)^3} \omega_1   S( \mathbf{q}- \mathbf{k}_1 )g(\mathbf{q}, \omega_1).
  \;
\end{split}
\end{equation}
Notice that due to the target's non-trivial polarization response, this form factor does not have a closed form. Thus, we numerically evaluate the decoherence effect, as stated in the main text.

\bibliography{ref}

\end{document}